\documentclass[journal]{IEEEtran}

\usepackage{mathrsfs}

\usepackage[noadjust]{cite}
\usepackage{graphicx,color,overpic,psfrag}
\usepackage{float}
\usepackage{epstopdf}
\usepackage{amsmath, amssymb}
\usepackage{cleveref}
\usepackage{latexsym}
\usepackage{bm}
\usepackage{amssymb}
\usepackage{cases}
\usepackage{array}
\usepackage{fancyhdr}
\usepackage{setspace}

\ifCLASSOPTIONcompsoc
\usepackage[caption=false,font=normalsize,labelfont=sf,textfont=sf]{subfig}
\else
\usepackage[caption=false,font=footnotesize]{subfig}
\fi

\definecolor{myblue}{rgb}{.91,.95,.99}
\usepackage{multirow}
\usepackage{url}
\usepackage{algorithm}
\usepackage{algorithmic}

\usepackage{blkarray}
\usepackage{booktabs}

\usepackage{multirow}
\usepackage{dsfont}
\usepackage{tabularx}
\usepackage[table]{xcolor}
\usepackage{gensymb}
\usepackage{amsfonts}

\usepackage{letltxmacro}

\graphicspath{{figure/}}

\newcommand{\ra}[1]{\renewcommand{\arraystretch}{#1}}
\newcolumntype{L}{>{\hspace*{-\tabcolsep}}l}
\newcolumntype{R}{c<{\hspace*{-\tabcolsep}}}
\definecolor{lightblue}{rgb}{0.93,0.95,1.0}

\newtheorem{remark}{Remark}

\newcommand{\figref}[1]{Fig. \ref{#1}}
\newcommand{\tabref}[1]{Table \ref{#1}}

\newcommand{\appref}[1]{Appendix \ref{#1}}
\newcommand{\secref}[1]{Section \ref{#1}}

\newcommand{\subfigref}[2]{Fig. \ref{#1}\subref{#2}}

\newcommand{\blkdiag}[1]{\mathsf{blkdiag}\left\{#1\right\}}

\newcommand{\abs}[1]{\left|#1\right|}

\newcommand{\thetabs}[2]{{\dnnot{\theta}{bs}}}

\newcommand{\be}{\mathbf{e}}

\newcommand{\bg}{\mathbf{g}}
\newcommand{\bh}{\mathbf{h}}

\newcommand{\bo}{\mathbf{o}}
\newcommand{\bp}{\mathbf{p}}
\newcommand{\bq}{\mathbf{q}}
\newcommand{\br}{\mathbf{r}}
\newcommand{\bs}{\mathbf{s}}

\newcommand{\bv}{\mathbf{v}}
\newcommand{\bw}{\mathbf{w}}
\newcommand{\bx}{\mathbf{x}}

\newcommand{\bA}{\mathbf{A}}
\newcommand{\bB}{\mathbf{B}}
\newcommand{\bC}{\mathbf{C}}
\newcommand{\bD}{\mathbf{D}}
\newcommand{\bE}{\mathbf{E}}

\newcommand{\bG}{\mathbf{G}}

\newcommand{\bI}{\mathbf{I}}
\newcommand{\bJ}{\mathbf{J}}

\newcommand{\bL}{\mathbf{L}}
\newcommand{\bM}{\mathbf{M}}

\newcommand{\bQ}{\mathbf{Q}}
\newcommand{\bR}{\mathbf{R}}

\newcommand{\bT}{\mathbf{T}}
\newcommand{\bU}{\mathbf{U}}
\newcommand{\bV}{\mathbf{V}}
\newcommand{\bW}{\mathbf{W}}
\newcommand{\bX}{\mathbf{X}}
\newcommand{\bY}{\mathbf{Y}}
\newcommand{\bZ}{\mathbf{Z}}

\newcommand{\bbC}{\mathbb{C}}

\newcommand{\dnnot}[2]{#1_{\mathrm{#2}}}

\allowdisplaybreaks

\begin{document}

\title{\huge Integrated Communications and Localization for Massive MIMO LEO Satellite Systems}

\author{
Li~You,
Xiaoyu~Qiang,
Yongxiang~Zhu,
Fan~Jiang,
Christos~G. Tsinos,
Wenjin~Wang,
Henk~Wymeersch,
Xiqi~Gao,
and~Bj\"{o}rn~Ottersten

\thanks{Copyright (c) 2015 IEEE. Personal use of this material is permitted. However, permission to use this material for any other purposes must be obtained from the IEEE by sending a request to pubs-permissions@ieee.org.}

\thanks{Part of this work was presented in ICC'2023 \cite{qiang2023hybrid}.}

\thanks{
Li You, Xiaoyu Qiang, Yongxiang Zhu, Wenjin Wang, and Xiqi Gao are with the National Mobile Communications Research Laboratory, Southeast University, Nanjing 210096, China, and also with the Purple
Mountain Laboratories, Nanjing 211100, China (e-mail: lyou@seu.edu.cn; xyqiang@seu.edu.cn; zhuyx@seu.edu.cn; wangwj@seu.edu.cn; xqgao@seu.edu.cn).
}
\thanks{Fan Jiang is with the Pengcheng Laboratory, Shenzhen 518000, China. He was with the School of Information Technology, Halmstad University, Halmstad 30118, Sweden (Email:  jiangf02@pcl.ac.cn)}
\thanks{
Christos G. Tsinos is with the National and Kapodistrian University of Athens, Evia, 34400, Greece and
also with the Interdisciplinary Centre for Security, Reliability
and Trust (SnT), University of Luxembourg, Luxembourg City 2721, Luxembourg (e-mail: ctsinos@uoa.gr).
}
\thanks{Henk Wymeersch is with the Department
of Electrical Engineering, Chalmers University of Technology, Gothenburg 41296, Sweden (e-mail: henkw@chalmers.se).
}
\thanks{
Bj\"{o}rn Ottersten is with the Interdisciplinary Centre for Security, Reliability
and Trust (SnT), University of Luxembourg, Luxembourg City 2721,
Luxembourg (bjorn.ottersten@uni.lu).
}
}
\maketitle

\begin{abstract}
Integrated communications and localization (ICAL) will play an important part in future sixth generation (6G) networks for the realization of
Internet of Everything (IoE) to support both global communications and seamless localization.
Massive multiple-input multiple-output (MIMO) low earth orbit (LEO) satellite systems have great potential in providing wide coverage with enhanced gains, and thus are strong candidates for realizing ubiquitous ICAL.
In this paper, we develop a wideband massive MIMO LEO satellite system to simultaneously support wireless communications and localization operations in the downlink.
In particular, we first characterize the signal propagation properties and derive a localization performance bound.
Based on these analyses, we focus on the hybrid analog/digital precoding design to achieve high communication capability and localization precision.
Numerical results demonstrate that the proposed ICAL scheme supports both the wireless communication and localization operations for typical system setups.
\end{abstract}


\section{Introduction}\label{sec:net_intro}
Fifth generation (5G) wireless networks are under deployment and the basic functionalities and capabilities are defined within the 5G standard \cite{you2021towards}.
However, there still exist many requirements that 5G networks may not satisfy, and sixth generation (6G) wireless networks are envisioned to offer seamless and ubiquitous coverage, higher communication capability and sensing/localization precision, and enhanced intelligence and security level, etc. \cite{de2021convergent,you2021towards,xiao2022overview,liu2022survey,ye2024fluid}.
One of the potential application scenarios of 6G networks is the integrated communications and localization (ICAL) on Internet of Everything (IoE), including tracking of persons or robots in an industrial site, autonomous driving, emergency response, etc.
In such use cases, communications and localization are simultaneously performed by by jointly designing the signal waveform for shared spectrum on one hardware platform, to improve the utilization of resources \cite{you2021towards,liu2022survey}.

One of the common application cases for ICAL on IoE is the terrestrial network \cite{jeong2014beamforming,jeong2018optimization,kwon2021joint}.
However, the ICAL functionality in a terrestrial network is unavailable in some areas where ground infrastructure is infeasible to deploy, or the signals are easily blocked \cite{li2022three}.
In these scenarios, satellite networks can provide an attractive and cost effective complement for the terrestrial networks since they can provide larger coverage, and support wideband communications and more flexible localization for the areas that terrestrial networks might have coverage issues.
Thus, satellite networks are expected to support global communications and seamless localization, and will play an essential role in performing ICAL for 6G networks \cite{liu2022survey,wang2021location,you2022hybrid,guidotti2022location}.
Generally, the satellite networks are divided into two categories, namely geostationary earth orbit (GEO) and non-GEO (NGEO) satellite networks \cite{al2021broadband}.
The typical satellite networks, including several global navigation satellite systems (GNSSs), e.g., global positioning system (GPS), GLONASS, and BEIDOU, are commonly capable of offering primary navigation with wide converge  \cite{li2022three}.
Those satellite networks are generally based on GEO and medium earth orbit (MEO) satellites, and recently, low earth orbit (LEO) satellite constellations have attracted much attention in terms of their application into position, navigation, and timing (PNT) \cite{prol2022position}.
The LEO satellites are usually deployed at altitudes of 200 -- 2000 km \cite{wang2019near}, and can be launched with low cost and high flexibility \cite{li2022three}.
Moreover, due to lower propagation delay and smaller path loss and footprint, the LEO satellite networks can provide better communication capability and localization precision compared with GEO counterparts \cite{li2022three,you2020massive}.
So far, several large LEO satellite systems, e.g., OneWeb, SpaceX, have been launched by governments and corporations, and see a steady reduction in launch costs,  which makes it possible to develop global LEO satellite systems, and complement GNSSs.

Massive multiple-input multiple-output (MIMO) can provide numerous degrees of freedom in both temporal and spatial domains \cite{liu2022survey}.
Besides, it can provide sufficient link budget to potentially support wideband communications to mobile terminals without dedicated antennas, and provide multiple links to do localization and tracking.
Therefore, it has gained much attention for pure communications and localization, to improve the spectral efficiency (SE) and the precision of localization \cite{you2020massive,shahmansoori2017position,mendrzik2018harnessing}, which motivates us to adopt the massive MIMO technology for ubiquitous ICAL \cite{jeong2018optimization,liu2022survey,xiao2022antenna}.
However, the implementation of fully digital transceivers in massive MIMO requires a large number of radio frequency (RF) chains, and might lead to high power consumption.
Generally, this issue can be circumvented by developing a hybrid precoding architecture \cite{el2014spatially}.
Recently, AST SpaceMobile has reported the successful deployment of the 693-square-foot MIMO array on its BlueWalker 3 LEO satellite \cite{rachel2022ast}.
Motivated by this, we combine the LEO satellite networks with the employment of massive MIMO in 6G, to support ICAL with the terrestrial user terminals (UTs) in the remote areas \cite{xiao2022overview}.

In this work, we propose to implement ICAL in the massive MIMO LEO satellite systems, to trade-off between the communication capability and the localization precision, which are evaluated by the SE and the squared position error bound (SPEB), respectively \cite{liu2022survey}.
Though the precoding designs for the downlink of the ICAL systems have been already investigated in the terrestrial networks \cite{jeong2014beamforming,jeong2018optimization,kwon2021joint}, the signal propagation properties in such systems differs from that of the LEO satellite ones, and thus can not be applied directly.
Specifically, owing to the mobility of the transceivers and the long satellite-to-UTs distance, there exist large Doppler shifts and a long propagation latency in the considered scenario \cite{you2020massive,xiao2022leo}.
Thus, the instantaneous channel state information (iCSI) between the satellite and the UTs is time-varying, which may be difficult to estimate. Moreover, the estimated iCSI might be outdated \cite{li2021downlink}, which
makes it challenging to use iCSI for downlink precoding in such system.
Motivated by these characteristics, we investigate the precoding design by exploiting the statistical CSI (sCSI), which is relatively slow-varying.

Inspired by the aforementioned motivations, a hybrid analog/digital transmitter is proposed for wideband massive MIMO LEO satellite systems to perform ubiquitous ICAL by exploiting sCSI.
The main contributions of the paper are summarized as follows:
\begin{itemize}
  \item We investigate the upper bound of the ergodic SE expression.
   Besides, we derive a closed-form Cram\'{e}r-Rao lower bound (CRLB) for the channel parameters of the considered systems, based on which the SPEB is derived to measure the performance of the downlink localization.
  \item We investigate the hybrid precoders multi-objective optimization for the considered systems, to trade-off between the communication capability and the localization precision, based on the SE and the SPEB metrics, respectively.
  \item We develop a hybrid precoding strategy and jointly design the signal waveform based on sCSI, to simultaneously perform communications and localization, and guarantee good performance in terms of both the SE as well as the SPEB metrics, respectively.
\end{itemize}
\subsection{Related Works}
\emph{Communications} -- So far, the communications for the LEO satellite scenarios have been intensively investigated.
In \cite{you2020massive}, the authors have formulated a massive MIMO communication scheme for both uplink and downlink of the LEO satellite systems based on the average signal-to-leakage-plus-noise ratio (ASLNR) and average signal-to-interference-plus-noise ratio (ASINR) maximization criteria, respectively.
The downlink precoding designs for both fully digital and hybrid transmitters have been studied in \cite{li2021downlink,you2022hybrid,you2022massive}, to maximize the downlink SE or the energy efficiency performance.
Besides, joint user scheduling and beamforming frameworks have been investigated for the downlink of the massive MIMO LEO satellite systems \cite{riviello2022joint,dakkak2023evaluation}.
In \cite{li2022uplink}, the authors focused on the research of the uplink transmit design for the massive MIMO LEO satellite systems.

\emph{Localization} -- Wireless localization can be performed with single anchor or multiple anchors, both of which have been extensively investigated in the terrestrial systems.
In \cite{shen2010fundamental1,shen2010fundamental2}, the authors have presented theoretical analyses for multiple anchor localizations.
The CRLB for single anchor localization has been derived for both two-dimensional (2D) and three-dimensional (3D) scenarios in \cite{shahmansoori2017position,abu2018error}.
In \cite{guerra2018single}, the authors have investigated the localization and orientation performance limits for the single anchor scenarios with massive MIMO transmission.
In \cite{kakkavas2019performance}, the authors have studied the influence of synchronization errors and Doppler effects on single anchor localization systems.

\emph{ICAL} -- The existing ICAL studies mainly focus on terrestrial scenarios.
In \cite{jeong2014beamforming,jeong2018optimization}, the authors have designed the beamforming vectors to simultaneously perform communications and localization during data transmission, based on rate maximization, SPEB minimization, or the transmission power minimization criteria.
Besides, localization can not only be performed together with the data transmission, but also with the pilot transmission, at the same time of channel estimation.
In \cite{kwon2021joint}, the authors have proposed a two-stage beamforming scheme, where in the first stage, pilot overhead signaling is minimized subject to localization precision constraints, and in the second stage, the data rate is maximized with the estimated CSI obtained from stage one.

\subsection{Organization}
The paper is organized as follows.
Section \ref{sec:sys_mod} formulates the system model for the wideband massive MIMO LEO satellite ICAL system.
The performance metrics for both communications and localization, i.e., SE and SPEB, are analyzed in Section \ref{sec:per_met}.
An algorithmic framework is developed in Section \ref{sec:ICAL} to design the hybrid analog/digital precoders for the ICAL system enabling the trade-off between the communication and localization performance.
Section \ref{sec:sim} presents the simulation results and the paper is concluded briefly in Section \ref{sec:conc}.

\subsection{Notations}
Matrices and vectors are denoted by upper and lower case boldface letters, respectively.
$\mathbb{C}^{m\times n}$ represents the $m\times n$-dimension unitary space.
The left-hand side of $\triangleq$ is defined by the right-hand side.
$\otimes$ denotes the Kronecker product.
$\exp\{\cdot\}$ and $\log\{\cdot\}$ are the exponential and logarithmic operators, respectively.
$\bI_{N}$ stands for $N\times N$ identity matrix.
$(\cdot)^T$, $(\cdot)^\ast$, and $(\cdot)^H$ represent the transpose, conjugate, and conjugate transpose operations, respectively.
$|x|$, $\angle x$, $\Re\left\{x\right\}$, and $\lceil x \rceil$ denote the amplitude, the angle, the real part, and the ceiling value of $x$, respectively.
The circular symmetric complex-valued zero-mean Gaussian distribution with variance $\sigma^2$ is given by $\mathcal{CN}(0,\sigma^2)$.
$\mathbb{E}\{\cdot\}$, ${\rm Tr}\left\{\cdot\right\}$, and $\blkdiag{}$ represent the expectation, the trace, and the block diagonal operators.
${\rm rank}\left\{\bX\right\}$ stands for the rank of the matrix $\bX$.
$||\cdot||_2$ and $||\cdot||_F$ denote the $\ell_2$-norm and Frobenius-norm, respectively.
The $(i,j)$th element of the matrix $\bA$ is given by $\left[\bA\right]_{i,j}$.
$\bA\succeq \bB$ refers to the positive semidefinite property of the matrix $\bA-\bB$.
$\partial$ denotes the partial derivative operation.
\section{System Model}\label{sec:sys_mod}
\begin{figure}[htbp]
		\centering
		\includegraphics[width=0.4\textwidth]{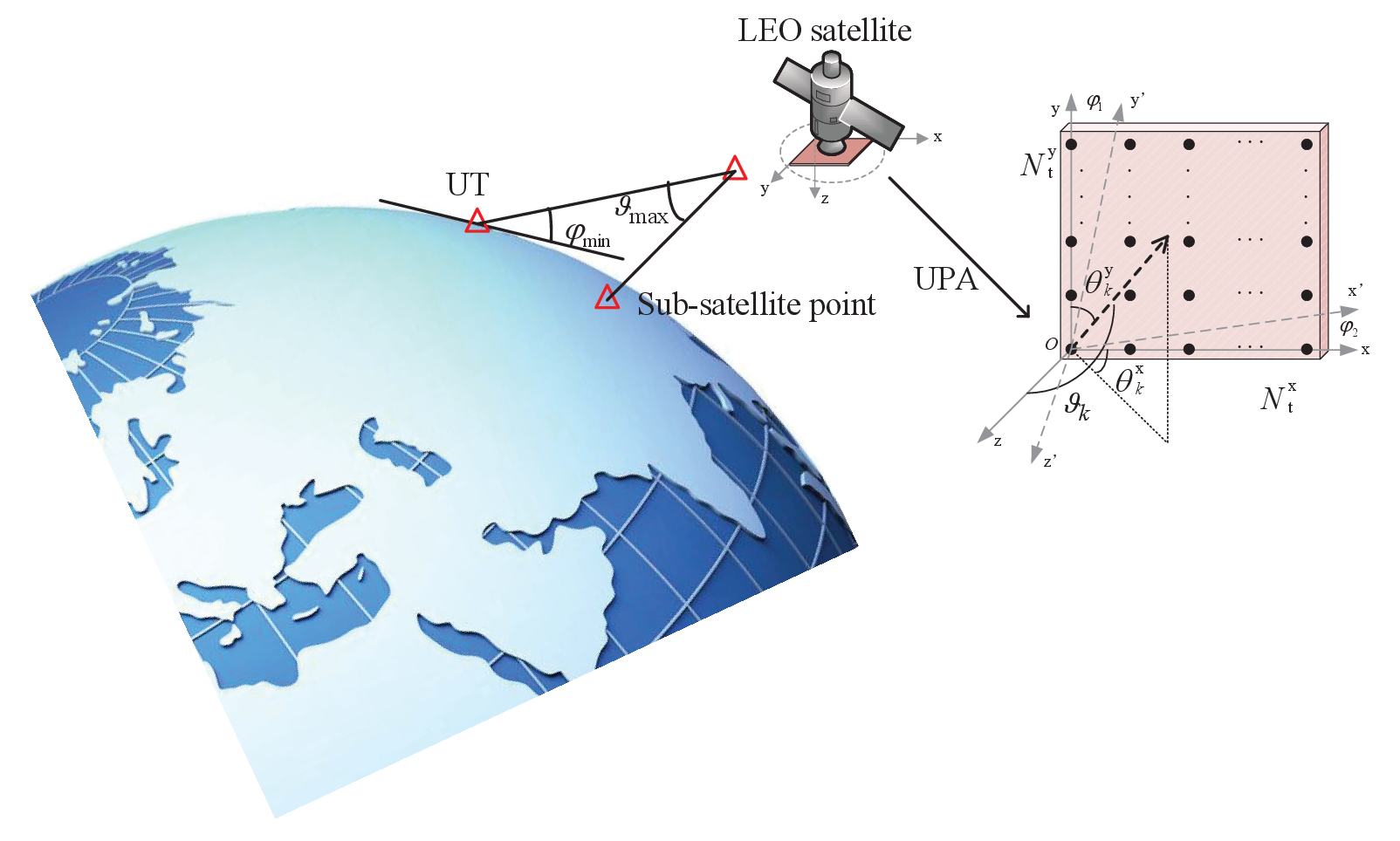}
		\caption{3D geometric model of the massive MIMO LEO satellite ICAL system with known satellite position and orientation, unknown UTs position.}
        \label{fig:localization}
	\end{figure}

We propose to simultaneously perform communications and localization for the massive MIMO LEO satellite systems, as depicted in \figref{fig:localization}.
The system is operated at carrier frequency $f_{\rm c}$ and the corresponding wavelength is given by $\lambda_{\rm c}=c/f_{\rm c}$, where $c$ denotes the speed of the light.
We assume $K$ single-antenna UTs, at unknown position $\bp_k=\left[p_k^{\rm x},p_k^{\rm y},p_k^{\rm z}\right]^T$ and velocity $\dot{\bp}_k=\left[\dot{p}_k^{\rm x},\dot{p}_k^{\rm y},\dot{p}_k^{\rm z}\right]^T$, $k=1,\ldots,K$, are served by a single LEO satellite with known position $\bq=\left[q^{\rm x},q^{\rm y},q^{\rm z}\right]^T$ and orientation angle $\bo=[\varphi_1,\varphi_2]^T$,\footnote{The orientation angle can be obtained and pre-compensated by, e.g., programmed tracking, accordingly with predicted movement of the LEO satellite \cite{maral2020satellite}.} where $\varphi_2$ and $\varphi_1$ refer to the rotation around positive ${\rm y}$- and negative ${\rm x'}$-axes,\footnote{After a rotation by $\varphi_2$ around positive ${\rm y}$-axis, the ${\rm y}$-coordinate does not change, i.e., ${\rm y'}={\rm y}$, while the ${\rm x}$- and ${\rm z}$-coordinates vary as ${\rm x'}={\rm z}\sin\varphi_2 +{\rm x}\cos\varphi_2$ and ${\rm z'}={\rm z}\cos\varphi_2 -{\rm x}\sin\varphi_2$, respectively.} respectively, as depicted in \figref{fig:localization}.
We assume fixed positions and velocities of the UTs over the observed interval and update them according to the large movements of the UTs. 
A uniform planar array (UPA) of $N_{\rm t}=N_{\rm t}^{\rm x}N_{\rm t}^{\rm y}$ antennas with half-wavelength separation is applied at the LEO satellite transmitter, where $N_{\rm t}^{\rm x}$ and $N_{\rm t}^{\rm y}$ denote the number of antennas at the ${\rm x}$- and ${\rm y}$-axes, respectively.
The satellite transmitter is supported by a hybrid precoder with $N_{\rm rf}$ ($K\leq N_{\rm rf}\leq N_{\rm t}$) RF chains.

\begin{figure}[!t]
		\centering
		\includegraphics[width=0.4\textwidth]{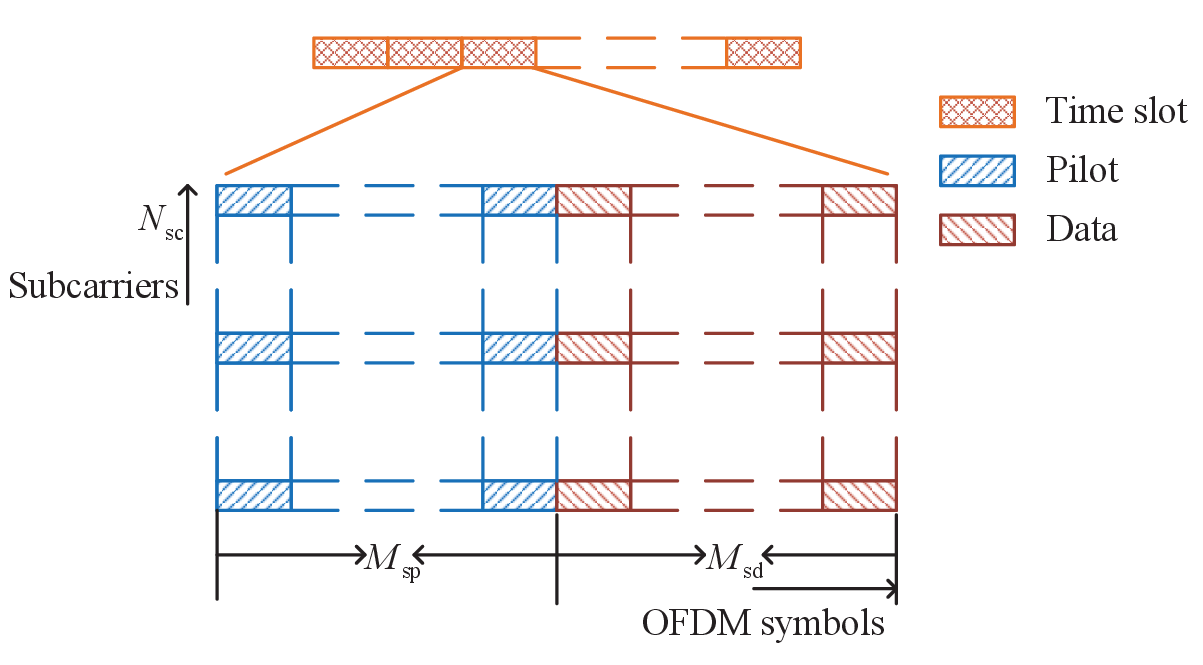}
		\caption{The time-frequency structure for the transmitted pilot and data signals.}
        \label{fig:frame}
	\end{figure}

The orthogonal frequency division multiplex (OFDM) modulation is employed for the downlink wideband transmission of the LEO satellite ICAL systems to mitigate the inter-symbol interference \cite{you2020massive,shahmansoori2017position}.
We denote $B_{\rm w}$ and $T_{\rm s}=1/(2B_{\rm w})$ as the system bandwidth and the sampling period, respectively.
In particular, we assume each frame consists of $M_{\rm s}$ slots, and there are $M_{\rm sp}$ and $M_{\rm sd}$ OFDM symbols used for pilot and data transmission in each slot, as depicted in \figref{fig:frame}.
Thus, in each frame, $M_{\rm p}=M_{\rm sp}M_{\rm s}$ and $M_{\rm d}=M_{\rm sd}M_{\rm s}$ OFDM symbols are transmitted through the pilot and data transmission, respectively.
Then, we assume $N_{\rm sc}$ subcarriers are employed over the system bandwidth $B_{\rm w}$, and the length of the cyclic prefix (CP) is set as $N_{\rm cp}$.
Thus, we denote $f_{\rm s}$ as the subcarrier separation and the frequency of the $n$th subcarrier is given by $f_n=(n-\frac{N_{\rm sc}+1}{2})f_{\rm s},\ n=1,\ldots,N_{\rm sc}$.
Subsequently, the OFDM symbol length with and without CP is given by $T=N_{\rm sc}T_{\rm s}+N_{\rm cp}T_{\rm s}$ and $T_{\rm sc}=N_{\rm sc}T_{\rm s}$, respectively.
In the following, let the subscript and superscript $g\in\left\{{\rm p},{\rm d}\right\}$ represent the pilot and data transmission, respectively.

An ICAL transmission protocol is developed for the LEO satellite systems.
First, rough position knowledge of both the satellite and the UTs can be obtained at the LEO satellite side by, e.g., initial access or tracking \cite{garcia2018optimal}, by exploiting which the satellite transmits precoded pilot and data signals to each UT.
Then, the required channel parameters can be evaluated from the received pilot signals at the UT, and more precise localization knowledge can be derived from the estimated channel information \cite{jeong2014beamforming,abu2018error,shahmansoori2017position}, to further improve the localization precision and communication capacity.
\subsection{Channel Model}\label{sec:chmd}
In the wideband  massive MIMO LEO satellite ICAL
systems, the UPA response is not only dependent on the angles-of-departure (AoD) information, but also the frequency.
Then, the UPA response $\bv_{k,l}\left(f\right)$ for the $l$th propagation path of the $k$th UT at frequency $f$ is given by \cite{you2022beam}
    \begin{align}\label{eq:chmdarr}
    \mathbf{v}_{k,l}\left(f\right)&=\mathbf{v}_{k,l}^{\mathrm{x}}\left(f\right)\otimes \mathbf{v}_{k,l}^{\mathrm{y}}\left(f\right)\notag\\
    &=\mathbf{v}_{\mathrm{x}}(f,\bm{\theta}_{k,l})\otimes\mathbf{v}_{\mathrm{y}}(f,\bm{\theta}_{k,l})\in \mathbb{C}^{N_{\rm t}\times 1},
    \end{align}
where $\bm{\theta}_{k,l}=\left(\theta_{k,l}^{\rm x},\theta_{k,l}^{\rm y}\right)$ denotes the AoD pair, as observed from \figref{fig:localization}.
Besides, the array response vectors $\mathbf{v}_{\mathrm{x}}(f,\bm{\theta}_{k,l})\in\mathbb{C}^{N_\mathrm{t}^{\rm x}\times 1}$ and $\mathbf{v}_{\mathrm{y}}(f,\bm{\theta}_{k,l})\in\mathbb{C}^{N_\mathrm{t}^{\rm y}\times 1}$ of the ${\rm x}$- and ${\rm y}$-axes can be expressed as \cite{you2020massive,you2022beam}
    \begin{align}\label{eq:crvc}
    \mathbf{v}_{k,l}^{\rm x}\left(f\right) &\triangleq \mathbf{v}_{\mathrm{x}}(f,\bm{\theta}_{k,l})
    =\frac{1}{\sqrt{N_\mathrm{t}^{\rm x}}}\left[1\ \exp\{-\jmath\varpi\sin{\theta_{k,l}^{\mathrm{y}}}\cos{\theta_{k,l}^{\mathrm{x}}}\}\ \right. \notag\\
    &\quad \quad \quad \left. \cdots\ \exp\{-\jmath\varpi(N_\mathrm{t}^{\rm x}-1)\sin{\theta_{k,l}^{\mathrm{y}}}\cos{\theta_{k,l}^{\mathrm{x}}}\}\right]^T,\\
    \mathbf{v}_{k,l}^{\rm y}\left(f\right) &\triangleq \mathbf{v}_{\mathrm{y}}(f,\bm{\theta}_{k,l})=\frac{1}{\sqrt{N_\mathrm{t}^{\rm y}}}\left[1\ \exp\{-\jmath\varpi\cos{\theta_{k,l}^{\mathrm{y}}}\}\ \cdots\ \right.\notag\\
     &\qquad \qquad \quad \quad \left. \exp\{-\jmath\varpi(N_\mathrm{t}^{\rm y}-1)\cos{\theta_{k,l}^{\mathrm{y}}}\}\right]^T,
    \end{align}
where $\varpi=\pi\left(1+f/f_{\rm c}\right)$.

Generally, the LEO satellite is deployed at an altitude much higher than that of the scatters around the UTs, then the AoD of each path for the channel associated with the $k$th UT is almost identical,\footnote{For an orbit height of about 200 km, the AOD difference of the x- and y-axes are about 0.03\degree and 0.01\degree when the scatterers are spread at a maximum radius of 100 m, which can be negligible.} i.e., $\bm{\theta}_{k,l}\triangleq\bm{\theta}_k,\ \forall l$, and thus we have $\bv_{k,l}\left(f\right)=\bv_k\left(f\right),\ \mathbf{v}_{k,l}^d\left(f\right)=\mathbf{v}_{k}^d\left(f\right)\in\mathbb{C}^{N_\mathrm{t}^d\times 1},\ \forall l,\ d\in\left\{{\rm x},{\rm y}\right\}$.
Let  $\bv_{k,n}\triangleq\bv_{k}\left(f_n\right)$, $\mathbf{v}_{k,n}^d\triangleq\mathbf{v}_k^d\left(f_n\right)$, and then, with  perfect time and frequency synchronization between the satellite and the UTs,\footnote{The clock bias/synchronization errors of the different UTs are not considered in the models and algorithms of this work.
In particular, synchronization can be assumed to be done by, e.g., a tracking algorithm or a joint localization and synchronization approach \cite{chen2021precoding} for simplicity (Also, the downlink transmission is a second phase of a real-time transport protocol, so that the distance can be determined by the time-of-arrival.).
Besides, in this work, we assume perfect carrier frequency offset synchronization between the UTs and the satellite, which can be obtained and then compensated by e.g., a under-sampling approach \cite{yuan2020hybrid}.
} the effective channel vector $\bh_{k,m_g,n}\in \bbC^{N_{\rm t}\times 1 }$ for the $k$th UT over the $n$th subcarrier of the $m_g$th OFDM symbol is given by $\bh_{k,m_g,n}=\bh^{\rm los}_{k,m_g,n}+\bh^{\rm nlos}_{k,m_g,n}$ \cite{li2021downlink},
where $\bh^{\rm los}_{k,m_g,n}$ and $\bh^{\rm nlos}_{k,m_g,n}$ denote the line-of-sight (LoS) and non-line-of-sight (NLoS) part of the channel, respectively, and they can be detailed as
\begin{align}
\bh^{\rm los}_{k,m,n}&=g^{\rm los}_{k,m_g,n}\bv_{k,n},\label{eq:losch}\\
\bh^{\rm nlos}_{k,m,n}&=g^{\rm nlos}_{k,m_g,n}\bv_{k,n}.\label{eq:nlosch}
\end{align}
Define $g_{k,m_g,n}=g^{\rm los}_{k,m_g,n}+g^{\rm nlos}_{k,m_g,n}$, where $g^{\rm los}_{k,m_g,n}$ and $g^{\rm nlos}_{k,m_g,n}$ denote the channel gains of the LoS and NLoS parts, respectively.
Then, since there are numerous propagation paths, $g_{k,m_g,n}$ can be approximated as the composition of a large number of independent and identically distributed components that follow the Rician distribution with Rician factor $\kappa_k$ and average power $\gamma_k=\mathbb{E}\{\abs{g_{k,m_g,n}}^2\}$ \cite{zhu2009mutual}.
In \eqref{eq:losch}, the complex gain $g^{\rm los}_{k,m_g,n}$ is given by $g^{\rm los}_{k,m_g,n}=\alpha_{k}\exp\left\{\jmath2\pi\left(\nu_{k}m_gT-nf_{\rm s}\tau_{k}\right)\right\}$,\footnote{Note that this model is valid for moderate Doppler spreads for which inter-carrier-interference can be  mitigated \cite{stamoulis2002intercarrier}.
}
where $\nu_{k}$ and $\tau_{k}$ are Doppler shifts and propagation delay of the LoS path with the $k$th UT, respectively.\footnote{The Doppler shift related to the mobility of the LEO satellite can be pre-compensated due to its deterministic time variation \cite{papathanassiou2001comparison}.}
Besides, $\alpha_{k}=\sqrt{\frac{\kappa_k\gamma_k}{1+\kappa_k}}\exp\{\jmath\phi_k\}$, where $\phi_k\in (0,2\pi]$ is a random phase.
In addition, in \eqref{eq:nlosch}, the complex gain $g^{\rm nlos}_{k,m_g,n}$ follows that $g^{\rm nlos}_{k,m,n}\sim \mathcal{CN}(0,\frac{\gamma_k}{1+\kappa_k})$.

\subsection{Signal Model}
We denote the transmitted pilot signal over the $m_{\rm p}$th OFDM symbol as $\{\bs_{m_{\rm p},n}^{\rm p}\}_{n=1}^{N_{\rm sc}^{\rm p}}$,
where $\bs_{m_{\rm p},n}^{\rm p}=\{s_{m_{\rm p},n,1}^{\rm p},\ldots,s_{m_{\rm p},n,K}^{\rm p}\},\ m_{\rm p}=1,\ldots,M_{\rm p}$ and satisfies
$\mathbb{E}\{\bs_{m_{\rm p},n}^{\rm p}(\bs_{m_{\rm p},n}^{\rm p})^H\}=\bI_K$.
Besides, the transmitted data signal over the $m_{\rm d}$th OFDM symbol is denoted as $\{\bs_{m_{\rm d},n}^{\rm d}\}_{n=1}^{N_{\rm sc}^{\rm d}}$,
where $\bs_{m_{\rm d},n}^{\rm d}=\{s_{m_{\rm d},n,1}^{\rm d},\ldots,s_{m_{\rm d},n,K}^{\rm d}\},\ m_{\rm d}=1,\ldots,M_{\rm d}$, $\mathbb{E}\{s_{m_{\rm d},n,k}^{\rm d}\}=0$ and $\mathbb{E}\{s_{m_{\rm d},n,k}^{\rm d}(s_{m'_{\rm d},n',k'}^{\rm d})^{\ast}\}=\delta(m_{\rm d}-m_{\rm d}')\delta(n-n')\delta(k-k')$ \cite{jeong2014beamforming}.
\begin{figure}[!b]
\centering
\subfloat[Fully connected.] {\includegraphics[width=0.4\textwidth]{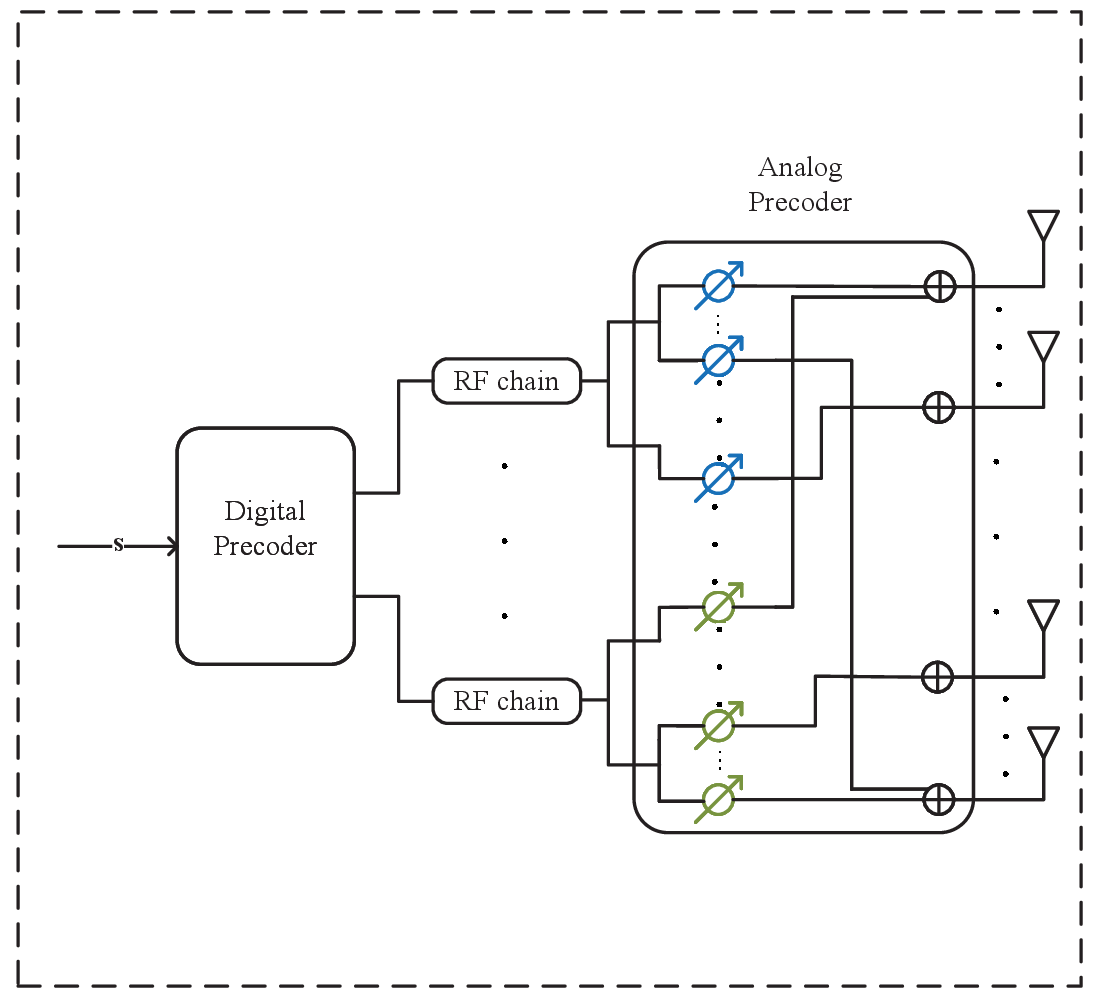}\label{HP_framework1}}
\hfill
\subfloat[Partially connected.] {\includegraphics[width=0.4\textwidth]{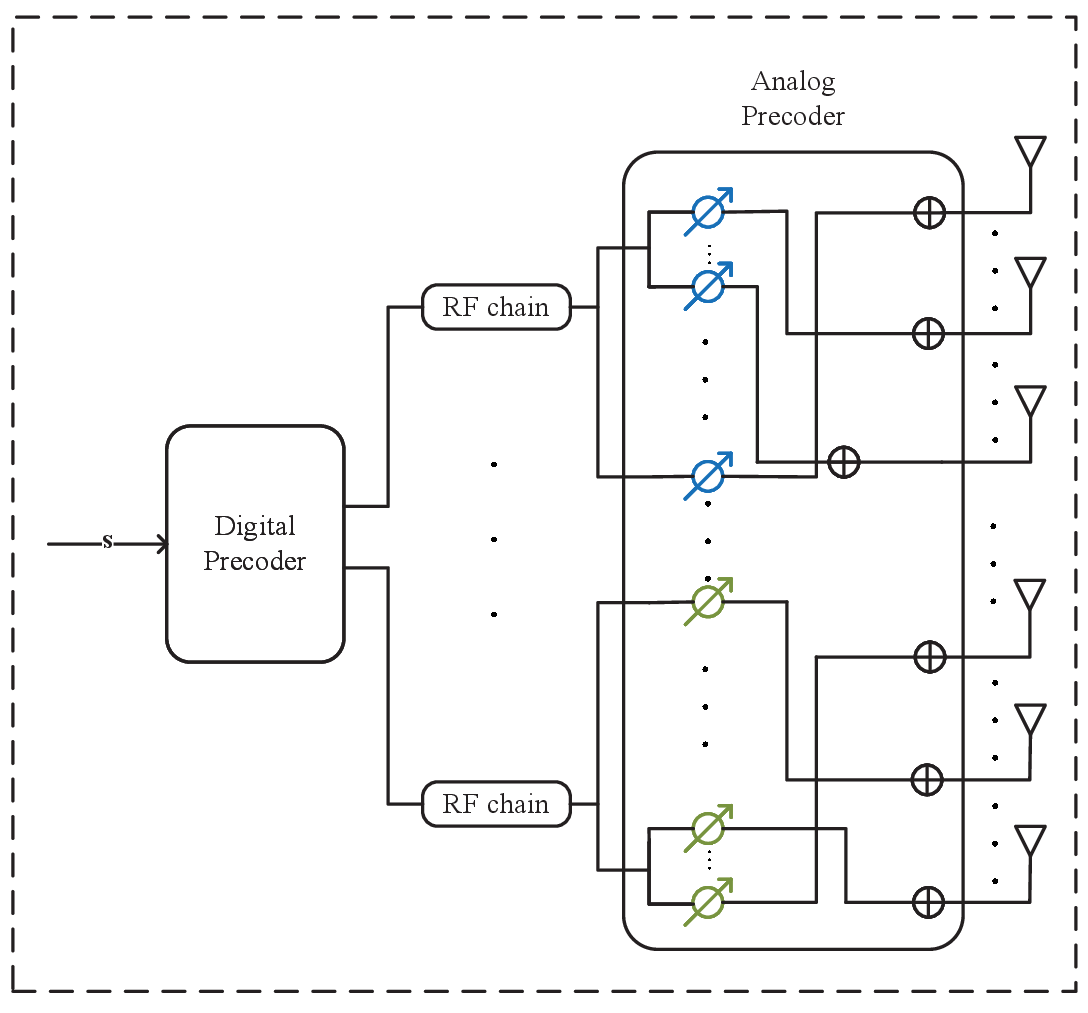}\label{HP_framework2}}
\caption{Hybrid precoding architectures of the massive MIMO LEO satellite ICAL system.}
\label{HP_framework}
\end{figure}

At the $n$th subcarrier over the $m_g$th OFDM symbol, the transmitted signal $\bs_{m_{g},n}^{g}$ is first processed by a baseband precoder $\bW_{{\rm BB},n}\in \mathbb{C}^{N_{\rm rf}\times K}$ and then by an analog precoder $\bW_{\rm RF}\in \mathbb{C}^{N_{\rm t}\times N_{\rm rf}}$ \cite{lin2020tensor}, as depicted in \figref{HP_framework}. Thus, the final transmission signal is given by $\bx_{m_{g},n}^{g}=\bW_{\rm RF}\bW_{{\rm BB},n}\bs_{m_{g},n}^{g}=\bW_{n}\bs_{m_{g},n}^{g}$,
where $\bW_{n}=\left[\bw_{n,1},\bw_{n,2},\ldots,\bw_{n,K}\right]$ is the equivalent hybrid precoding matrix.
Then, the received pilot/data signal over the $n$th subcarrier of the $m_g$th OFDM symbol at the $k$th UT is given by
\begin{align}
y_{k,m_g,n}^g&=\bh_{k,m_g,n}^T\bW_{n}\bs_{m_g,n}^{g}+z_{k,m_g,n}^g,
\end{align}
where $z_{k,m_{g},n}^{g}\sim \mathcal{CN}\left(0,N_0\right),\ \forall k,m_{g},n,\ \forall g\in\left\{{\rm p},{\rm d}\right\}$.

\section{Performance Metrics}\label{sec:per_met}
\subsection{Communication Spectral Efficiency}
In the following, we first omit the OFDM symbol and subcarrier indices $m_{\rm d}$ and $n$ for a clearer description of the channel statistical properties.
During data transmission, the ergodic data rate for the $k$th UT is given by
    \begin{align}\label{eq:rateput1}
    R_k=\mathbb{E}\left\{\log\left(1 + \frac{|\mathbf{h}_{k}^T\bw_{k}|^2}{\sum_{\ell\neq k}|\mathbf{h}_{k}^T\bw_{\ell}|^2+N_0}\right)\right\}.
    \end{align}
Note that Monte Carlo method can be used to estimate the ergodic date rate.
However, the computational complexity of the Monte Carlo method is extremely high, and the accurate iCSI is difficult to be obtained at the LEO satellite transmitter, as mentioned in the introduction, which motivates us to adopt sCSI. 
Based on the \cite[Lemma 2]{sun2015beam}, it can be concluded that the logarithmic expression inside the expectation operator of $R_k$ is concave with respect to the matrix $\bh_k^{\ast}\bh_k^T$ and thus, is upper bounded by
\begin{align}\label{eq:jiub}
R_k\leq \bar{R}_k=\log\left(1+\frac{\bw_k^H\mathbb{E}\left\{\bh_k^{\ast}\bh_k^T\right\}\bw_k}{\sum_{\ell\neq k}\bw_{\ell}^H\mathbb{E}\left\{\bh_k^{\ast}\bh_k^T\right\}\bw_{\ell}}\right).
\end{align}

In \eqref{eq:jiub}, the expectation expression $\mathbb{E}\left\{\bh_k^{\ast}\bh_k^T\right\}$ represents the channel correlation matrix at the satellite side for the $k$th UT.
Thus, we have $\mathbb{E}\left\{\bh_k^{\ast}\bh_k^T\right\}=\bar{\bh}_k^{\ast}\bar{\bh}_k^T$ and
\begin{align}\label{eq:eqch}
\bar{\bh}_k=\sqrt{\gamma_k}\bv_k.
\end{align}
Note that the required sCSI knowledge involves the channel gain $\gamma_k$ and the UPA response $\bv_k$, which are regarded to be constant during the observed interval and can be updated dynamically in accordance with the channel variation \cite{you2020massive}.
Subsequently, the upper bound of the ergodic rate with the sCSI knowledge is given by
\begin{align}\label{eq:rateub}
\bar{R}_k
=\log\left(1 + \frac{|\bar{\mathbf{h}}_{k}^T\bw_{k}|^2}{\sum_{\ell\neq k}|\bar{\mathbf{h}}_{k}^T\bw_{\ell}|^2+N_0}\right),
\end{align}
where $\bar{\bh}_k$ is given in \eqref{eq:eqch}.
The tightness of the upper bound has been established for the Rician channel assumption in \cite{matthaiou2009ergodic}, where the upper bound is proved to be even tighter with smaller transmission power and larger Rician factor.
Besides, the tightness will also be verified by simulations in \secref{sec:sim}.
Finally, we can express the SE as\footnote{For notation brevity, the OFDM symbol index $m_{\rm d}$ is omitted here as the following focus is on each time slot.}
\begin{align}
R_{\rm sum}&=\frac{1}{B_{\rm w}}\sum_{k=1}^K\sum_{n=1}^{N_{\rm sc}}f_{\rm s}\bar{R}_{k,n}\notag\\
&=\frac{1}{B_{\rm w}}\sum_{k=1}^K\sum_{n=1}^{N_{\rm sc}}f_{\rm s}\log\left(1 + \frac{|\bar{\mathbf{h}}_{k,n}^T\bw_{k,n}|^2}{\sum_{\ell\neq k}|\bar{\mathbf{h}}_{k,n}^T\bw_{\ell,n}|^2+N_0}\right),
\end{align}
where $\bar{\mathbf{h}}_{k,n}=\sqrt{\gamma_{k,n}}\bv_{k,n}$.

\subsection{Localization Accuracy}
\subsubsection{Received Signal Model}
The received pilot signal at the $k$th UT is given by\footnote{The superscript and subscript ${\rm p}$ is omitted for notation brevity.}
\begin{align}\label{eq:rsig}
y_{k,m,n}\notag
=\left(\left(\bh_{k,m,n}^{\rm los}\right)^T+\left(\bh_{k,m,n}^{\rm nlos}\right)^T\right)\bW_{n}\bs_{m,n}+z_{k,m,n}.
\end{align}
For notation convenience, we denote $\tilde{z}_{k,m,n}=\left(\bh_{k,m,n}^{\rm nlos}\right)^T\bW_{n}\bs_{m,n}+z_{k,m,n}$, and then, \eqref{eq:rsig} can be converted into $y_{k,m,n}=\left(\bh_{k,m,n}^{\rm los}\right)^T\bW_{n}\bs_{m,n}+\tilde{z}_{k,m,n}$,
where $\tilde{z}_{k,m,n}$ follows $\mathcal{CN}(0,N_{k,n}^{\rm eq})$, and the variance
$N_{k,n}^{\rm eq}$ is given by $N_{k,n}^{\rm eq}=\mathbb{E}\{\abs{\tilde{z}_{k,m,n}}^2\}=\frac{\gamma_k}{1+\kappa_k}\bv_{k,n}^T\bW_n\bW_n^H\bv_{k,n}^{\ast}+N_0$.
\subsubsection{Fisher Information Matrix (FIM)}
As mentioned before, channel parameters can be estimated from the received signals.
The corresponding channel parameters between the satellite and the $k$th UT can be characterized by a $6\times 1$ vector
$\bm{\eta}_k=[\theta_{k}^{\rm x},\theta_{k}^{\rm y},\tau_{k},\nu_{k},\alpha_{k}^{\rm R},\alpha_{k}^{\rm I}]^T$,
where $\alpha_{k}^{\rm R}$ and $\alpha_{k}^{\rm I}$ are the real and imaginary parts of $\alpha_{k}$, respectively.
Then, we denote $\hat{\bm{\eta}}_k$ as an estimate for the parameter vector $\bm{\eta}_k$ associated with the $k$th UT, the mean squared error (MSE) of which allows a lower bound, given by $\mathbb{E}\{(\hat{\bm{\eta}}_k-\bm{\eta}_k)(\hat{\bm{\eta}}_k-\bm{\eta}_k)^T\}\succeq \bJ_{\bm{\eta}_k}^{-1}$.
Note that $\bJ_{\bm{\eta}_k}$ is the FIM for the unknown vector $\bm{\eta}_k$ and the $(i,j)$th element can be computed from \cite{steven1993fundamentals,mendrzik2018harnessing,kakkavas2019performance,zhang2021using}
\begin{align}
\left[\bJ_{\bm{\eta}_k}\right]_{i,j}&=\sum_{m=1}^{M}\sum_{n=1}^{N_{\rm sc}}\left[\bJ_{\bm{\eta}_k}\right]_{i,j}^{m,n}\notag\\
&=\sum_{m=1}^{M}\sum_{n=1}^{N_{\rm sc}}\frac{2}{N_{k,n}^{\rm eq}}\Re\left\{\mathbb{E}\left\{\frac{\partial \left(r_{k,m,n}\right)^H}{\partial \left[\bm{\eta}_k\right]_i}\frac{\partial r_{k,m,n}}{\partial \left[\bm{\eta}_k\right]_j}\right\}\right\},
\end{align}
where $r_{k,m,n}=(\bh_{k,m,n}^{\rm los})^T\bW_n\bs_{m,n}$ is the received pilot signal excluding the noise.
\subsubsection{Transformation for Position Parameters}
Note that the transformation from channel parameters $\bm{\eta}_{k}$ to $\bar{\bm{\eta}}_{k}=[\bp_k^T,\alpha_{k}^{\rm R},\alpha_{k}^{\rm I}]^T$ is a bijection.
Then, the transformed FIM involved the position information associated with the $k$th UT is given by $\bJ_{\bar{\bm{\eta}}_{k}}=\bm{\Gamma}_{k}\bJ_{\bm{\eta}_{k}}\bm{\Gamma}_{k}^T$.
The transformation matrix $\bm{\Gamma}_{k}\in \mathbb{R}^{5\times 6}$ is given by
\begin{align}\label{eq:trmtr}
\bm{\Gamma}_{k}\triangleq \frac{\partial \bm{\eta}_{k}^T}{\partial \bar{\bm{\eta}}_{k}}={\rm blkdiag}\left\{\bm{\Xi}_{k},\bI_{2}\right\},
\end{align}
where $\bm{\Xi}_{k}$ is detailed in \appref{app:b}.
\subsubsection{SPEB}
Localization precision can be measured by the metric squared position error (SPE), the definition of which is the MSE between the actual position $\bp_k$ of the $k$th UT and its estimation $\hat{\bp}_k$ \cite{jeong2014beamforming},
i.e., $\rho_k\left(\bW\right)=\mathbb{E}\left\{\abs{\abs{\hat{\bp}_k-\bp_k}}_2^2\right\}$.
Following the information inequality, the bound of the SPE is given by $\rho_k\left(\bW\right)\geq {\rm Tr}\left\{\left(\bJ^{\rm e}_{\bp_k}\right)^{-1}\right\}$
\cite{shen2010fundamental1}.
Then, we define the sum SPEB of the UTs as
\begin{align}
\rho_{\rm sum}^{\rm b}=\sum_{k=1}^K{\rm Tr}\left\{\left[\bJ_{\bar{\bm{\eta}}_{k}}^{-1}\right]_{1:3,1:3}\right\}=\sum_{k=1}^K{\rm Tr}\left\{\bE^T\bJ_{\bar{\bm{\eta}}_k}^{-1}\bE\right\}.
\end{align}
where $\bE=[\mathbf{e}_1,\mathbf{e}_2,\mathbf{e}_3]$, and $\mathbf{e}_i\in \mathbb{R}^{5\times 1}$ denotes a vector with the $i$th element being one while the others being zero.
\section{Integrated Communications and Localization}\label{sec:ICAL}
\subsection{Problem Formulation}\label{sec:prof}
Our work aims to design a hybrid precoding approach for the considered LEO satellite ICAL systems.
To that end, an optimization problem is formulated, to maximize the downlink SE\footnote{Here, we omit the constant $f_{\rm s}/B_{\rm w}$ in the expression of the SE for brevity.} and minimize the sum SPEB.
Let $\mathcal{W}=\{\bW_{{\rm BB},n}\}_{n=1}^{N_{\rm sc}}$, and we define a vector of the objectives as $f(\mathcal{W},\bW_{\rm RF})=[R_{\rm sum}, -\rho_{\rm sum}^{\rm b}]^T$, then the multi-objective optimization problem is given by \cite{marler2004survey}
    \begin{subequations}\label{eq:srpmax}
    \begin{align}
    \mathcal{P}_1:\mathop{\mathrm{maximize}}\limits_{\mathcal{W},\bW_{\rm RF}}&\ \ f\left(\mathcal{W},\bW_{\rm RF}\right) \label{eq:srpmaxa}\\
    \mathrm{s.t.}&\ \ \sum_{n=1}^{N_{\rm sc}}||\bW_{\rm RF}\bW_{{\rm BB},n}||_F^2\leq P, \label{eq:srpmaxb}\\
    &\ \ \bW_{\rm RF}\in \mathcal{S}.\label{eq:srpmaxc}
    \end{align}
    \end{subequations}
Note that problem $\mathcal{P}_1$ is the maximization problem of the vector $f(\mathcal{W},\bW_{\rm RF})$ with both communication and localization metrics, which is defined to simultaneously maximize those two elements \cite{bjornson2014multiobjective}.
In Eq. \eqref{eq:srpmax}, $P$ is the transmit power budget, and $\mathcal{S}\in \{\mathcal{S}_{\rm FC},\mathcal{S}_{\rm PC}\}$ for the fully and partially connected structures, respectively.
For the partially connected structure, the antennas are divided in to $N_{\rm rf}$ groups, where each group allows $N_{\rm g}=N_{\rm t}/N_{\rm rf}$ antenna elements.
Thus, the corresponding analog precoding matrix is block diagonal, given by $\bW_{\rm RF}=\blkdiag{\bw_{\rm RF,1},\ldots,\bw_{\rm RF,N_{\rm rf}}}$.
Besides, the communication and localization metric components of the objective in Eq. \eqref{eq:srpmaxa} have different units, and thus, we respectively transform them into dimensionless ones in the first place \cite{marler2004survey}.
Then, a weighted sum method is exploited for the problem with transformed objectives to trade-off between the performance of communications and localization, where a positive weighting coefficient is selected to satisfy the condition for Pareto optimality \cite[Definition 1]{marler2004survey}.
Note that several systematic techniques have been developed to determine the weighting coefficient, i.e., ranking methods \cite{yoon1995multiple} and eigenvalue method \cite{saaty1977scaling}.
Finally, we jointly design the digital precoder for each subcarrier and the analog precoder for all subcarriers.
\subsection{Spectral Efficiency Maxmization}\label{sec:srmax}
For notation brevity, let the product of the hybrid precoders denoted by $\bW_n=\bW_{\rm RF}\bW_{{\rm BB},n}$.
By utilizing weighted minimum MSE (WMMSE) method, the maximization of SE can be equivalently transformed into minimizing the weighted sum MSE \cite{shi2011iteratively}.
In particular, we assume that linear combiner $u_{k,n}$ is incorporated at subcarrier $n$ of the $k$th UT.
Then, by letting $\bU=\{u_{k,n}\}_{k=1,n=1}^{K,N_{\rm sc}}$ and introducing an auxiliary weight variable $\bm{\Omega}=\{\omega_{k,n}\}_{k=1,n=1}^{K,N_{\rm sc}}$, the weighted sum MSE is given by $\varepsilon^{\rm sum}=\sum_{k=1}^K\sum_{n=1}^{N_{\rm sc}}\omega_{k,n}\varepsilon_{k,n}\left(\bW_n,u_{k,n}\right)-{\rm log}\left(\omega_{k,n}\right)$,
where $\varepsilon_{k,n}(\bW_n,u_{k,n})$ is the MSE between the estimated signal $\hat{s}_{k,n}=u_{k,n}^{\ast}y_{k,n}$ and the transmitted signal $s_{k,n}$, given by
\begin{align}
&\varepsilon_{k,n}\left(\bW_n,u_{k,n}\right)=\mathbb{E}\left\{\left(\hat{s}_{k,n}-s_{k,n}\right)\left(\hat{s}_{k,n}-s_{k,n}\right)^H\right\}\notag\\
=&|u_{k,n}^{\ast}\bar{\bh}_{k,n}^T\bw_{k,n}-1|^2+\sum_{i\neq k}|u_{k,n}^{\ast}\bar{\bh}_{k,n}^T\bw_{i,n}|^2+N_0|u_{k,n}|^2.
\end{align}
Note that when the transmitter precoders are known, the minimum MSE (MMSE) receiver can be given by $u_{k,n}^{\rm MMSE}=\bar{\bh}_{k,n}^T\bw_{k,n}(\sum_{i=1}^K|\bar{\bh}_{k,n}^T\bw_{i,n}|^2+N_0)^{-1}$.
\subsection{Sum SPEB Minimization}\label{sec:sbmin}
Note that the sum SPEB component of the objective in Eq. \eqref{eq:srpmaxa} is not easy to tackle due to the tightly coupled digital and analog precoding matrices.
In this section, we regard the product of the digital and analog precoding matrices as a whole, i.e., $\bW_n^{\rm loc}=\bW_{\rm RF}\bW_{{\rm BB},n},\ \forall n$, and our objective is to find the fully digital precoder $\bW_n^{\rm loc}$ to minimize the sum SPEB, which can be formulated as
    \begin{align}\label{eq:spebmin1}
    \mathcal{P}_2:\mathop{\mathrm{minimize}}\limits_{\{\bW_n^{\rm loc}\}_{n=1}^{N_{\rm sc}}}& \sum_{k=1}^K{\rm Tr}\left\{\bE^T\bJ_{\bar{\bm{\eta}}_k}^{-1}\bE\right\} \
    \mathrm{s.t.}\ \ \sum_{n=1}^{N_{\rm sc}}||\bW_n^{\rm loc}||_F^2\leq P.
    \end{align}
Note that problem $\mathcal{P}_2$ has been investigated in the terrestrial systems, where the elements of FIM are linearly dependent on the optimization variable. However, for the LEO satellite scenarios, the linear dependence no longer exists, and problem $\mathcal{P}_2$ requires transformation into a convex problem to be effectively addressed.
The minimization of the sum SPEB is converted into minimizing the sum Euclidean distance between the product of the hybrid precoders and the precoders obtained from problem $\mathcal{P}_2$.
Then, let $\bC_n\triangleq\bW_n^{\rm loc}(\bW_n^{\rm loc})^H$ and problem $\mathcal{P}_2$ can be converted into a rank-constrained problem
   \begin{subequations}\label{eq:spebmin2}
    \begin{align}
    \mathcal{P}_3:\mathop{\mathrm{minimize}}\limits_{\bC_n}&\sum_{k=1}^K{\rm Tr}\left\{\bE^T\bJ_{\bar{\bm{\eta}}_k}^{-1}\bE\right\} \label{eq:irmax2a}\\
    \mathrm{s.t.}&\sum_{n=1}^{N_{\rm sc}}{\rm Tr}\left\{\bC_n\right\}\leq P,\ \bC_n\succeq \bm{0},\ {\rm rank}\left\{\bC_n\right\}\leq K.\label{eq:spebmin2d}
    \end{align}
    \end{subequations}
Then, we introduce an auxiliary variable $\bM_k\in \mathbb{R}^{3\times 3}$, which satisfies the following condition \cite{li2013robust}
\begin{align}\label{eq:intov}
\bM_k\succeq \bE^T\bJ_{\bar{\bm{\eta}}_k}^{-1}\bE, \ \forall k.
\end{align}
Note that the FIM $\bJ_{\bar{\bm{\eta}}_k}$ must be a positive semidefinite matrix, and thus, utilizing the property of Schur complement, Eq. \eqref{eq:intov} can be transformed into \cite{li2013robust}
\begin{align}\label{eq:intovsc}
\begin{bmatrix}
\bM_k &\bE^T\\
\bE &\bJ_{\bar{\bm{\eta}}_k}\left(\bC_n\right)
\end{bmatrix}\succeq \bm{0}, \ \forall k.
\end{align}
Therefore, problem $\mathcal{P}_3$ can be converted into \cite{li2013robust,keskin2022optimal}
    \begin{subequations}\label{eq:spebmin3}
    \begin{align}
    \mathcal{P}_4:\mathop{\mathrm{minimize}}\limits_{\bC_n,\bM_k}&\ \ \sum_{k=1}^K{\rm Tr}\left\{\bM_k\right\} \label{eq:irmax3a}\\
    \mathrm{s.t.}&\ \ \sum_{n=1}^{N_{\rm sc}}{\rm Tr}\left\{\bC_n\right\}\leq P,\label{eq:spebmin3b}\\
    &\ \ \bC_n\succeq \bm{0},\label{eq:spebmin3c}\\
    &\ \ {\rm rank}\left\{\bC_n\right\}\leq K,\label{eq:spebmin3d}\\
    &\ \
\begin{bmatrix}
\bM_k &\bE^T\\
\bE &\bJ_{\bar{\bm{\eta}}_k}\left(\bC_n\right)
\end{bmatrix}\succeq \bm{0}, \ \forall k. \label{eq:spebmin3e}
    \end{align}
    \end{subequations}
Note that the rank constraint in \eqref{eq:spebmin3d} of problem $\mathcal{P}_4$ is difficult to handle.
Therefore, we first focus on the relaxed problem by dropping this rank constraint \cite{boyd2004convex}.
To further improve computational efficiency, based on \cite[Appendix C]{li2007range}, the relaxed problem of $\mathcal{P}_4$ allows a optimal solution, which is given by
\begin{align}\label{eq:optC}
\bC_n=\bL_n\bZ_n\bL_n^H.
\end{align}
The proof is given in \appref{app:g}.
In Eq. \eqref{eq:optC}, $\bZ_n$ denotes a $3K\times 3K$-dimensional positive semidefinite matrix and $\bL_n=\left[\bV_n\ \bV_{n,{\rm x}}\ \bV_{n,{\rm y}}\right]$,
where $\bV_n=\left[\bv_{1,n},\ldots,\bv_{K,n}\right]$, $\bV_{n,d}=\left[\bv_{1,n,\theta_1^d},\ldots,\bv_{K,n,\theta_K^d}\right]$, and $\bv_{k,n,\theta_1^d}\triangleq\partial\bv_{k,n}/\partial \theta_1^d$ for $d\in \left\{{\rm x},{\rm y}\right\}$.
By utilizing the decomposition in Eq. \eqref{eq:optC}, problem $\mathcal{P}_4$ can be converted into
    \begin{subequations}\label{eq:spebmin4}
    \begin{align}
    \mathcal{P}_5:\mathop{\mathrm{minimize}}\limits_{\bZ_n,\bM_k}&\ \ \sum_{k=1}^K{\rm Tr}\left\{\bM_k\right\} \label{eq:irmax4a}\\
    \mathrm{s.t.}&\ \ \sum_{n=1}^{N_{\rm sc}}{\rm Tr}\left\{\bL_n\bZ_n\bL_n^H\right\}\leq P,\label{eq:spebmin4b}\\
    &\ \ \bZ_n\succeq \bm{0},\label{eq:spebmin4c}\\
    &\ \
\begin{bmatrix}
\bM_k &\bE^T\\
\bE &\bJ_{\bar{\bm{\eta}}_k}\left(\bZ_n\right)
\end{bmatrix}\succeq \bm{0}, \ \forall k. \label{eq:spebmin4d}
    \end{align}
    \end{subequations}
To tackle problem $\mathcal{P}_5$, since $\bJ_{\bar{\bm{\eta}}_k}(\bZ_n)$ is not convex with respect to the variable $\bZ_n$, we invoke the majorization-minimization (MM) algorithm, the basic philosophy of which is to iteratively handle problem $\mathcal{P}_5$ through a series of easier problems \cite{sun2016majorization}.
In particular, let $\bZ_{n,t}$ denote the solution of the $t$th iteration, and then, in $(t+1)$th iteration, we substitute $\bJ_{\bar{\bm{\eta}}_k}(\bZ_n)$ with its
second order Taylor expansion $\hat{\bJ}_{\bar{\bm{\eta}}_k}(\bZ_n)$, whose $\left(i,j\right)$th element is given in \eqref{eq:sote} on the top of the next page.
In \eqref{eq:sote}, the first order gradient of $\left[\bJ_{\bar{\bm{\eta}}_k}\right]_{i,j}^{m,n}\left(\bZ_{n,t}\right)$ with respect to the variable $\bZ_{n,t}$ is given in \eqref{eq:pofz} on the top of the next page,
where $\bA_{k,m,n,i,j}=\bL_n^H\frac{\partial \bh_{k,m,n}^{\ast}}{\partial \left[\bm{\eta}_k\right]_i}\frac{\partial \bh_{k,m,n}^T}{\partial \left[\bm{\eta}_k\right]_j}\bL_n$ and $\bD_{k,n}=\frac{\gamma_k}{1+\kappa_k}\bL_n^H\bv_{k,n}^{\ast}\bv_{k,n}^T\bL_n$.
\newcounter{TempEqCnt1} 
\setcounter{TempEqCnt1}{\value{equation}} 
\setcounter{equation}{23} 
\begin{figure*}[!t] 
    \begin{align}\label{eq:sote}
    \left[\hat{\bJ}_{\bar{\bm{\eta}}_k}\right]_{i,j}\left(\bZ_n\right)= \left[\bJ_{\bar{\bm{\eta}}_k}\right]_{i,j}\left(\bZ_{n,t}\right)+\sum_{m=1}^{M}\sum_{n=1}^{N_{\rm sc}} {\rm Tr}\left\{\left(\frac{\partial \left[\bJ_{\bar{\bm{\eta}}_k}\right]_{i,j}^{m,n}\left(\bZ_{n,t}\right)}{\partial \bZ_{n,t}}\right)^T\left(\bZ_n-\bZ_{n,t}\right)\right\}
    +\frac{L}{2}\abs{\abs{\bZ_n-\bZ_{n,t}}}_F^2
    \end{align}
    \hrule
\end{figure*}
\newcounter{TempEqCnt2} 
\setcounter{TempEqCnt2}{\value{equation}} 
\setcounter{equation}{24} 
\begin{figure*}[!t] 
    \begin{align}\label{eq:pofz}
    \frac{\partial \left[\bJ_{\bar{\bm{\eta}}_k}\right]_{i,j}^{m,n}\left(\bZ_{n,t}\right)}{\partial \bZ_{n,t}}=\frac{\left({\rm Tr}\left\{\bD_{k,n}\bZ_{n,t}\right\}+N_0\right)\bA_{k,m,n,i,j}-{\rm Tr}\left\{\bZ_{n,t}\bA_{k,m,n,i,j}\right\}\cdot\bD_{k,n}}{\left({\rm Tr}\left\{\bD_{k,n}\bZ_{n,t}\right\}+N_0\right)^2}
    \end{align}
    \hrule
\end{figure*}

Then, in the $\left(t+1\right)$th iteration, the corresponding problem can be written as
    \begin{subequations}\label{eq:spebmin4t}
    \begin{align}
    \mathcal{P}_5^{\left(t+1\right)}:\mathop{\mathrm{minimize}}\limits_{\bZ_{n,t+1},\bM_{k,t+1}}&\ \ \sum_{k=1}^K{\rm Tr}\left\{\bM_{k,t+1}\right\} \label{eq:irmax4ta}\\
    \mathrm{s.t.}&\ \ \sum_{n=1}^{N_{\rm sc}}{\rm Tr}\left\{\bL_n\bZ_{n,t+1}\bL_n^H\right\}\leq P,\label{eq:spebmin4tb}\\
    &\ \ \bZ_{n,t+1}\succeq \bm{0},\label{eq:spebmin4tc}\\
    &\ \
\begin{bmatrix}
\bM_k &\bE^T\\
\bE &\hat{\bJ}_{\bar{\bm{\eta}}_k}\left(\bZ_{n,t+1}\right)
\end{bmatrix}\succeq \bm{0}, \ \forall k, \label{eq:spebmin4td}
    \end{align}
    \end{subequations}
which can be handled with semidefinite programming (SDP) solvers \cite{vandenberghe1996semidefinite,kakkavas20195g}.
Note that since the MM method results in a relaxed problem, each stationary point of the series involving the objective values produced by problem $\mathcal{P}_5^{\left(t+1\right)}$ might be a local sub-optimum of problem $\mathcal{P}_5$ \cite{jacobson2007}.
Then, with the symmetric positive definite matrix $\bZ_n$ from $\mathcal{P}_5$, we have $\bC_n=\bL_n\bZ_n\bL_n^H$, and the corresponding localization precoder can be reversed through Cholesky decomposition and randomization procedures \cite{golub2013matrix,luo2010semidefinite,kishore2017literature,garcia2018optimal}.

\remark Due to the roughness of the prior knowledge obtained at the LEO satellite side for the $k$th UT, robust signal design should be formulated with the consideration of the uncertainties in channel parameters, which can be performed by referring to \cite{garcia2018optimal,keskin2022optimal}.
\subsection{Hybrid Precoding for the ICAL Systems}
Let $\bar{\bW}_{\rm BB}=[\bW_{{\rm BB},1},\ldots,\bW_{{\rm BB},N_{\rm sc}}]\in \mathbb{C}^{N_{\rm rf}\times KN_{\rm sc}}$ and $\bar{\bW}^{\rm loc}=[\bW_1^{\rm loc},\ldots,\bW_{N_{\rm sc}}^{\rm loc}]\in \mathbb{C}^{N_{\rm t}\times N_{\rm rf}N_{\rm sc}}$ for notation brevity.
To design the digital and analog precoders, the following weighted sum problem is formulated to minimize the sum MSE of communications and the sum Euclidean distance between the hybrid precoders and the localization-only fully digital precoders, as obtained in \secref{sec:srmax} and \secref{sec:sbmin}, which is given by \cite{wan2021hybrid,cheng2021hybrid,el2014spatially}
    \begin{subequations}\label{eq:wscl1}
    \begin{align}
    \mathcal{Q}_1:&\mathop{\mathrm{minimize}}\limits_{\substack{\bW_{\rm RF},\{\mathbf{W}_{{\rm BB},n}\}_{n}^{N_{\rm sc}},\\ \bU,\bm{\Omega}}}\ \ \rho\varepsilon^{\rm sum}\left(\bW_{\rm RF},\bar{\bW}_{\rm BB},\bU,\bm{\Omega}\right)\notag\\
    &\qquad \qquad \quad \quad \quad +(1-\rho)d^{\rm sum}\left(\bW_{\rm RF},\bar{\bW}_{\rm BB}\right) \label{eq:wscl1a}\\
    &\qquad \qquad \ \ \mathrm{s.t.}\qquad ||\bW_{\rm RF}\bar{\bW}_{\rm BB}||_F^2\leq P,\ \bW_{\rm RF}\in \mathcal{S}. \label{eq:wscl1c}
    \end{align}
    \end{subequations}
In \eqref{eq:wscl1}, $\rho\in [0,1]$ denotes the weighting coefficient, serving to trade off between the performance of the communications and localization, and $d^{\rm sum}$ is defined as $d^{\rm sum}(\bW_{\rm RF},\bar{\bW}_{\rm BB})=\sum_{n=1}^{N_{\rm sc}}\abs{\abs{\bW_{\rm RF}\bW_{{\rm BB},n}-\bW_n^{\rm loc}}}_F^2$.
By introducing the auxiliary variables $\bG_{k,n}=\left[\bg_{k,n}^1,\ldots,\bg_{k,n}^{K}\right]=\bW_{\rm RF}\bW_{{\rm BB},n},\ \forall k$, problem $\mathcal{Q}_1$ can be reformulated as \cite{cheng2021hybrid}
    \begin{subequations}\label{eq:wscl2}
    \begin{align}
    \mathcal{Q}_2:&\mathop{\mathrm{minimize}}\limits_{\substack{\bW_{\rm RF},\{\mathbf{W}_{{\rm BB},n}\}_{n=1}^{N_{\rm sc}},\\ \bU,\bm{\Omega}}}\ \ \sum_{k=1}^K\sum_{n=1}^{N_{\rm sc}}\Big[\rho\varepsilon_{k,n}^{\rm w}\left(\bG_{k,n},u_{k,n},\omega_{k,n}\right)\notag\\
    &\qquad \qquad \qquad \qquad \qquad \quad +\frac{1-\rho}{K}d_{k,n}\left(\bG_{k,n}\right)\Big] \label{eq:wscl2a}\\
    &\qquad \qquad \ \ \mathrm{s.t.}\qquad \bG_{k,n}=\bW_{\rm RF}\bW_{{\rm BB},n},\ \forall k,\label{eq:wscl2b}\\
    &\qquad \qquad \qquad \qquad \sum_{k=1}^K\sum_{n=1}^{N_{\rm sc}}{\rm Tr}\left\{\bG_{k,n}\bG_{k,n}^H\right\}\leq PK, \notag\\
    &\qquad \qquad \qquad \qquad \qquad \quad \bW_{\rm RF}\in \mathcal{S}, \label{eq:wscl2d}
    \end{align}
    \end{subequations}
where $\varepsilon_{k,n}^{\rm w}(\bG_{k,n},u_{k,n},\omega_{k,n})=\omega_{k,n}\varepsilon_{k,n}(\bG_{k,n},u_{k,n})-{\rm log}(\omega_{k,n})$, $d_{k,n}=\abs{\abs{\bG_{k,n}-\bW_n^{\rm loc}}}_F^2$ and
$\varepsilon_{k,n}(\bG_{k,n},u_{k,n})=|u_{k,n}|^2\bar{\bh}_{k,n}^T\bG_{k,n}\bG_{k,n}^H\bar{\bh}_{k,n}^{\ast}-u_{k,n}\bar{\bh}_{k,n}^T\bG_{k,n}\be_k-u_{k,n}^{\ast}\bG_{k,n}^H\bar{\bh}_{k,n}^{\ast}\be_k^T+1+N_0|u_{k,n}|^2$.

In the following, inspired by the  alternating direction method of multipliers (ADMM) \cite{boyd2011distributed}, we adopt the augmented Lagrangian method by introducing dual variables $\bQ_{k,n}\in \mathbb{C}^{N_{\rm t}\times K}$, and the corresponding penalty $\eta_{k,n}> 0$.
Then, let $\mathcal{G}=\left\{\bG_{k,n}\right\}_{k=1,n=1}^{K,N_{\rm sc}}$, $\mathcal{Q}=\left\{\bQ_{k,n}\right\}_{k=1,n=1}^{K,N_{\rm sc}}$ and the objective of problem $\mathcal{Q}_2$ can be transformed into
\begin{align}
&f\left(\mathcal{G},\bU,\bm{\Omega},\bar{\mathbf{W}}_{\rm BB},\bW_{\rm RF},\mathcal{Q}\right)\notag\\
=&\sum_{k=1}^K\sum_{n=1}^{N_{\rm sc}}L\left(\bG_{k,n},u_{k,n},\omega_{k,n},\mathbf{W}_{{\rm BB},n},\bW_{\rm RF},\bQ_{k,n}\right),
\end{align}
where $L\left(\bG_{k,n},u_{k,n},\omega_{k,n},\mathbf{W}_{{\rm BB},n},\bW_{\rm RF},\bQ_{k,n}\right)=\rho\varepsilon_{k,n}^{\rm w}\left(\bG_{k,n},u_{k,n},\omega_{k,n},\bQ_{k,n}\right)+\frac{1-\rho}{K}d_{k,n}\left(\bG_{k,n}\right)+\frac{\eta_{k,n}}{2}\abs{\abs{\bG_{k,n}-\bW_{\rm RF}\mathbf{W}_{{\rm BB},n}+\bQ_{k,n}}}_F^2$.
The corresponding problem $\mathcal{Q}_3$ can be then written as
    \begin{subequations}\label{eq:wscl3}
    \begin{align}
    \mathcal{Q}_3:&\mathop{\mathrm{minimize}}\limits_{\substack{\mathcal{G},\bU,\bm{\Omega},\\ \bar{\mathbf{W}}_{\rm BB},\bW_{\rm RF},\mathcal{Q}}}\ \ f\left(\mathcal{G},\bU,\bm{\Omega},\bar{\mathbf{W}}_{\rm BB},\bW_{\rm RF},\mathcal{Q}\right) \label{eq:wscl3a}\\
    &\qquad \quad \ \mathrm{s.t.}\quad \sum_{k=1}^K\sum_{n=1}^{N_{\rm sc}}{\rm Tr}\left\{\bG_{k,n}\bG_{k,n}^H\right\}\leq PK, \label{eq:wscl3b}\\
    &\qquad \qquad \qquad \bW_{\rm RF}\in \mathcal{S}, \label{eq:wscl3c}
    \end{align}
    \end{subequations}
which can be handled through the ADMM method \cite{boyd2011distributed}. In each iteration of the ADMM method, the algorithmic steps mainly depend on the optimization of each variable in problem $\mathcal{Q}_3$.
It is worth noting that, in each iteration, the architecture of the ADMM allows us to update variables $\bG_{k,n}$, $u_{k,n}$, $\omega_{k,n}$, and $\bW_{{\rm BB},n}$ in parallel \cite{bertsekas1997nonlinear}.
Then, the corresponding steps for handling problem $\mathcal{Q}_3$ by the ADMM are given by
\begin{align}
\mathcal{G}^{t}&\leftarrow \arg \min_{\eqref{eq:wscl3b}} f\left(\mathcal{G},\bU^{t-1},\bm{\Omega}^{t-1},\bar{\mathbf{W}}_{\rm BB}^{t-1},\bW_{\rm RF}^{t-1},\mathcal{Q}^{t-1}\right),\label{eq:bcd1}\\
\mathcal{U}^{t}&\leftarrow \arg \min f\left(\mathcal{G}^{t},\bU,\bm{\Omega}^{t-1},\bar{\mathbf{W}}_{\rm BB}^{t-1},\bW_{\rm RF}^{t-1},\mathcal{Q}^{t-1}\right),\label{eq:bcd2}\\
\bm{\Omega}^{t}&\leftarrow \arg \min f\left(\mathcal{G}^{t},\bU^{t},\bm{\Omega},\bar{\mathbf{W}}_{\rm BB}^{t-1},\bW_{\rm RF}^{t-1},\mathcal{Q}^{t-1}\right),\label{eq:bcd3}\\
\bar{\mathbf{W}}_{\rm BB}^{t}&\leftarrow \arg \min f\left(\mathcal{G}^{t},\bU^{t},\bm{\Omega}^{t},\bar{\mathbf{W}}_{\rm BB},\bW_{\rm RF}^{t-1},\mathcal{Q}^{t-1}\right),\label{eq:bcd4}\\
\bW_{\rm RF}^{t}&\leftarrow \arg \min_{\eqref{eq:wscl3c}}  f\left(\mathcal{G}^{t},\bU^{t},\bm{\Omega}^{t},\bar{\mathbf{W}}_{\rm BB}^{t},\bW_{\rm RF},\mathcal{Q}^{t-1}\right),\label{eq:bcd5}\\
\bQ_{k,n}^{t}&\leftarrow\bQ_{k,n}^{t-1}+\bG_{k,n}^{t}-\bW_{\rm RF}^{t}\mathbf{W}_{{\rm BB},n}^{t}\label{eq:bcd6},
\end{align}
the solution of which is detailed in \appref{app:a}.
To further improve the convergence rate, at the end of each iteration, the penalty $\eta_{k,n}$ is updated according to \cite[Eq. (3.13)]{boyd2011distributed}
\begin{align}\label{eq:updss}
\eta_{k,n}^{t+1}=
\left\{
  \begin{array}{ll}
    \varsigma^{\rm mul}\eta_{k,n}^{t}, & \frac{\abs{\abs{\bG_{k,n}^{t+1}-\bW_{\rm RF}^{t+1}\mathbf{W}_{{\rm BB},n}^{t+1}}}_F^2}{\abs{\abs{\bG_{k,n}^{t+1}-\bG_{k,n}^{t}}}_F^2}>\varrho, \\
\eta_{k,n}^{t}/\varsigma^{\rm div}, & \frac{\abs{\abs{\bG_{k,n}^{t+1}-\bW_{\rm RF}^{t+1}\mathbf{W}_{{\rm BB},n}^{t+1}}}_F^2}{\abs{\abs{\bG_{k,n}^{t+1}-\bG_{k,n}^{t}}}_F^2}<1/\varrho, \\
    \eta_{k,n}^{t}, & \text{otherwise},
  \end{array}
\right.
\end{align}
where $\varrho>1$ and $\varsigma^{\rm mul},\varsigma^{\rm div}>1$.
The whole procedure for computing the hybrid precoders for LEO satellite ICAL is summarized in \textbf{Algorithm \ref{alg:algdasam}}.

\begin{algorithm}[!t]
\caption{Hybrid Precoding for ICAL}
\label{alg:algdasam}
\begin{algorithmic}[1]
\REQUIRE Thresholds $\epsilon_1,\epsilon_2>0$.
\ENSURE Hybrid precoders $\bW_{\rm RF}$ and $\bW_{{\rm BB},n}$, $n=1,\ldots,N_{\rm sc}$.
\STATE Initialize $t=0$, $t_{\rm max}^{\rm loc}$ and $\bZ_{n,t}$ such that $||\bL_n\bZ_{n,t}\bL_n||_F^2=P/N_{\rm sc}$.
\REPEAT
\STATE Solve SDP problem $\mathcal{P}_8^{\left(t+1\right)}$ and obtain $\bM_{k,t+1}$, $\bZ_{n,t+1}$.
\STATE $t=t+1$.
\UNTIL{$\abs{\sum_{k=1}^K{\rm Tr}\left\{\bM_{k,t}\right\}-\sum_{k=1}^K{\rm Tr}\left\{\bM_{k,t-1}\right\}}>\epsilon_1$ or $t\geq t_{\rm max}^{\rm loc}$.}
\STATE Obtain $\bC_n=\bL_n\bZ_n\bL_n^H$ and $\bW_{n}^{\rm loc}$ by eigenvalue decomposing $\bC_n$, $n=1,\ldots,N_{\rm sc}$.
\STATE Initialize $t=0$, $\mathcal{G}^t,\bU^t,\bm{\Omega}^t,\bar{\mathbf{W}}_{\rm BB}^t,\bW_{\rm RF}^t,\mathcal{Q}^t$, $t_{\rm max}^{\rm hy}$.
\WHILE{$\sum_{k=1}^K\sum_{n=1}^{N_{\rm sc}}\abs{\abs{\bG_{k,n}^t-\bW_{\rm RF}^t\bW_{{\rm BB},n}^t}}_F^2>\epsilon_2$ or $t\leq t_{\rm max}^{\rm hy}$}
\STATE $t=t+1$.
\STATE Update $\mathcal{G}^{t}$, $\mathcal{U}^{t}$, $\bm{\Omega}^{t}$, $\bar{\mathbf{W}}_{\rm BB}^{t}$, $\bW_{\rm RF}^{t}$, $\bQ_{k,n}^{t}$ and $\eta_{k,n}^t$ following Eqs. \eqref{eq:bcd1} -- \eqref{eq:updss}.
\ENDWHILE
\end{algorithmic}
\end{algorithm}
\subsection{Convergence and Computational Complexity}
The convergence of \textbf{Algorithm \ref{alg:algdasam}} depends on two parts.
First, the sum SPEB minimization problem is handled with MM method by transforming the projection matrices, i.e., ${\rm Tr}\left\{\bE^T\bJ_{\bar{\eta}_k}^{-1}\bE\right\}$, with their second-order Taylor expansion, and according to \cite{sun2016majorization}, the sequence of feasible points $\left\{\bZ_{n,t}\right\}$ can converge to a stationary value.
According to \cite{hong2016convergence}, the ADMM algorithm is guaranteed to be convergent to a stationary point when the penalty is properly chosen.

The complexity of \textbf{Algorithm \ref{alg:algdasam}} is detailed as follows.
First, in each iteration of the MM algorithm, the number of the optimized variables for the SDP in \eqref{eq:spebmin4t} is denoted by $n_{\rm var}=9K^2N_{\rm sc}+9K$.
The number of linear matrix inequality (LMI) constraints is $M_{\rm LMI}=K+N_{\rm sc}$, contributed by \eqref{eq:spebmin4tc} and \eqref{eq:spebmin4td}, respectively.
Besides, the number of the rows or columns for the matrix of the $i$th LMI constraint is given by $m_i=9, \ 1\leq i\leq K$ and $m_i=3K, \ K+1\leq i\leq M_{\rm LMI}$.
Thus, the complexity of the SDP in \eqref{eq:spebmin4t} is given by $\mathcal{O}(n^2_{\rm var}\sum_{i=1}^{M_{\rm LMI}}m_i^2+n_{\rm var}\sum_{i=1}^{M_{\rm LMI}}m_i^3)$ \cite{keskin2022optimal}.
By assuming the MM algorithm terminates in $J_{\rm mm}$ iterations, the overall complexity to handle problem $\mathcal{P}_5$ can be evaluated as $\mathcal{O}(J_{\rm mm}(n^2_{\rm var}\sum_{i=1}^{M_{\rm LMI}}m_i^2+n_{\rm var}\sum_{i=1}^{M_{\rm LMI}}m_i^3))$.
The computational complexity of the eigenvalue decomposition for $\bC_n,\ n=1,\ldots,N_{\rm sc}$ is given by $\mathcal{O}(N_{\rm sc}N_{\rm t}^3)$.
Subsequently, in each iteration of the ADMM, the corresponding parameters should be updated according to Eqs. \eqref{eq:bcd1} -- \eqref{eq:updss}, and the major computational complexity lies in the update of $\bG_{k,n}$.
In particular, the update for $\bG_{k,n}$ in step Eq. \eqref{eq:bcd1} mainly contributes to the pseudo-inverse operation and the eigenvalue decomposition, both of which present the complexity of $\mathcal{O}(N_{\rm t}^3)$.
The computation of $u_{k,n}$ and $\omega_{k,n}$ in Eqs. \eqref{eq:bcd2} and \eqref{eq:bcd3} mainly depends on the multiplication operation $\bar{\bh}_{k,n}^T\bG_{k,n}$, which requires the complexity of $\mathcal{O}(N_{\rm t}K)$.
Besides, both updates for the baseband precoder $\bW_{{\rm BB},n}$ and the analog precoder $\bW_{\rm rf}$ in Eqs. \eqref{eq:bcd4} and \eqref{eq:bcd5} contributes $\mathcal{O}(N_{\rm rf}^3)$ to the algorithm's complexity.
In addition, the computational complexity of both Eqs. \eqref{eq:bcd6} and \eqref{eq:updss} results from the multiplication of $\bW_{\rm RF}$ and $\bW_{{\rm BB},n}$, which can be evaluated as $\mathcal{O}(N_{\rm t}N_{\rm rf}K)$.
Then, assuming the ADMM algorithm terminates in $J_{\rm admm}$ iterations, the complexity for the ADMM can be estimated as $\mathcal{O}(2J_{\rm admm}KN_{\rm sc}N_{\rm t}^3)$.
\section{Simulations}\label{sec:sim}
In this section, we evaluate the communication and localization performance for the proposed massive MIMO LEO satellite ICAL system, which is assumed to operate in the S-band.
The performance of the communications and localization is evaluated with the downlink SE and average position error bound (APEB), which is defined as $\rho^{\rm b}_{\rm avg}=\sqrt{\rho^{\rm b}_{\rm sum}/K}$, respectively.
Some related typical simulation parameters are given in \tabref{tab:test} \cite{38.821,38.211}.
The LEO satellite is assumed to be located at $\bq=[0,0,0]^T$ and the orientation angle is given by $\bo=[0,0]^T$.
The maximum nadir angle of the UT is set to be $\vartheta_{\rm max}=\pi/6$ and each 
component of the AoD pair for the UT, i.e., $\theta_{k}^{\rm x}$ and $\theta_k^{\rm y}$, is assumed to be uniformly distributed in $\left[\pi/2-\vartheta_{\rm max},\pi/2+\vartheta_{\rm max}\right]$.
The nadir angle of the $k$th UT can be calculated as $\vartheta_k=\arccos\left(\sin\theta_{k}^{\rm x}\sin\theta_{k}^{\rm y}\right)$ and thus, the elevation angle for the $k$th UT is given by $\varphi_k=\arccos\left(\frac{R_{\rm e}+H}{R_{\rm e}}\sin\vartheta_k\right)$, where $R_{\rm e}$ is the earth radius and $H$ is the orbit height of the LEO satellite \cite{li2021downlink,lutz2012satellite}.
Then, the distance between the LEO satellite and the $k$th UT can be calculated as
$d_k=\sqrt{H^2+2HR_{\rm e}+R_{\rm e}^2\sin^2\varphi_k}-R_{\rm e}\sin\varphi_k$.
The channel gain $\gamma_{k}$ is defined as
\begin{align}\label{eq:normchp}
\gamma_{k} = G_{\rm sat}G_{\rm ut}N_\mathrm{t}\left(\frac{c}{4\pi f_{\rm c}d_k}\right)^2,
\end{align}
where $G_{\rm sat}$ and $G_{\rm ut}$ denote the antenna gain at the satellite and the UTs, respectively.
The velocity of each UT at ${\rm x}$, ${\rm y}$, or ${\rm z}$-axes is assumed to be uniformly distributed in [-10,10] m/s.
\begin{table*}[!t]
\caption{Simulation Parameters}\label{tab:test}
\centering
\ra{1.3}
\footnotesize
\begin{tabular}{LLR}
\toprule
&\ Parameter &\ Value  \\
\midrule
\multirow{11}{*}{Channel} &\cellcolor{lightblue}System bandwidth $B_{\rm w}$ & \cellcolor{lightblue} 15.36 MHz\\
&Carrier frequency $f_c$ &\ 2 GHz \\
&\cellcolor{lightblue}Speed of light $c$ &\cellcolor{lightblue} 3$\times 10^8$ m/s\\
&Carrier wavelength $\lambda_c$ &\ 0.15 m \\
&\cellcolor{lightblue}Subcarrier separation $f_{\rm s}$ &\cellcolor{lightblue}  30 kHz\\
&Number of valid subcarriers $N_{\rm sc}$ &\ 512 \\
&\cellcolor{lightblue}Sampling rate $1/T_{\rm s}$ &\cellcolor{lightblue}  $30.72$ MHz\\
&Number of slots per frame $M_{\rm s}$ &\ 20 \\
&\cellcolor{lightblue}Number of CP $N_{\rm cp}$ & \cellcolor{lightblue} 36\\
&Rician factor $\kappa_k$ &\ 18 dB\\
&\cellcolor{lightblue}Number of OFDM symbols per slot $M_{\rm sp}$, $M_{\rm sd}$ &\cellcolor{lightblue}  2, 12\\
\midrule
\multirow{5}{*}{Satellite} & \cellcolor{lightblue}Orbit height $H$ &\cellcolor{lightblue} 200 km\\
&Number of antennas $N_{\rm t}$ & 24$\times$24\\
&\cellcolor{lightblue}Antenna spacing $r_{\rm x}$, $r_{\rm y}$ &\cellcolor{lightblue} $\lambda_c/2$ \\
&Number of RF chains $N_{\rm rf}$ &\ 36 \\
&\cellcolor{lightblue}Antenna gain $G_{\rm sat}$, $G_{\rm ut}$ & \cellcolor{lightblue} 6 dB, 0 dB\\
\midrule
\multirow{2}{*}{UTs} &\cellcolor{lightblue}Number of UTs $K$ &\cellcolor{lightblue} 9\\
&Noise spectral density &\ -174 dBm/Hz\\
\bottomrule
\end{tabular}
\end{table*}


%
\begin{figure}[!t]
\centering
\subfloat[SE.]{\includegraphics[width=0.42\textwidth]{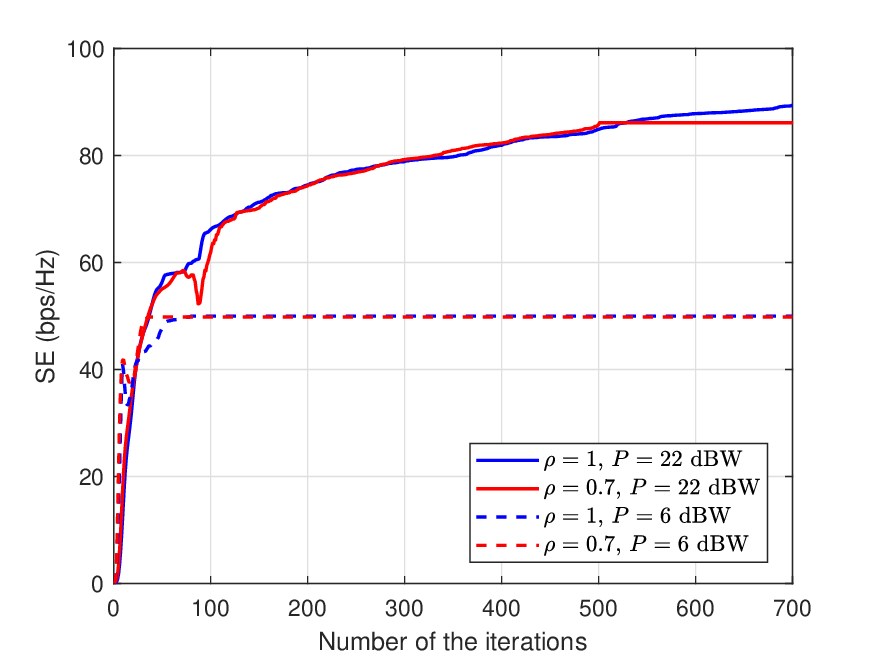}\label{se_conver}}
\hfill
\subfloat[APEB.]{\includegraphics[width=0.42\textwidth]{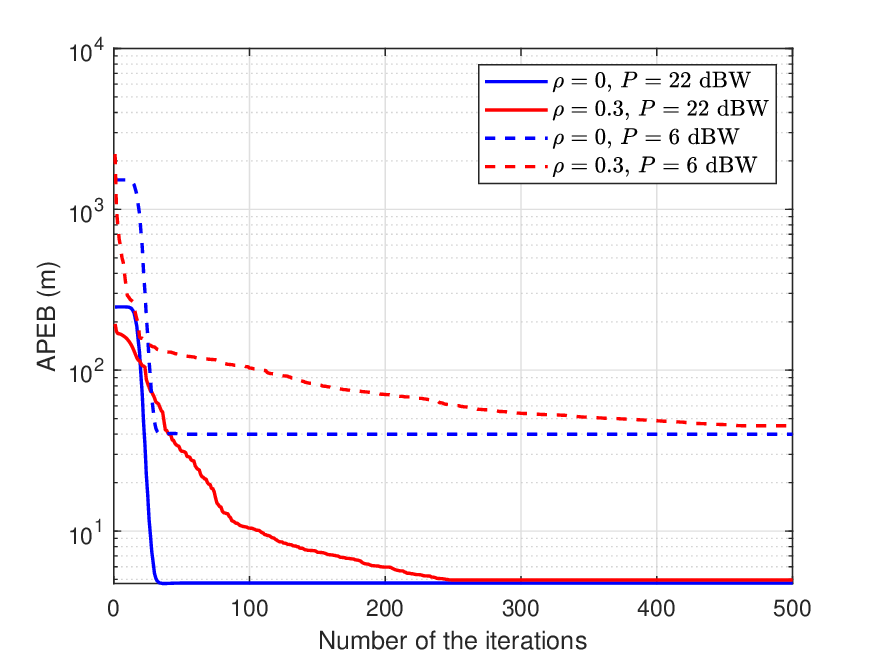}\label{speb_conver}}
\caption{The SE and APEB versus the number of iterations.}
\label{fig_conver}
\end{figure}

\begin{figure}[!t]
\centering
\subfloat[SE.]{\includegraphics[width=0.4\textwidth]{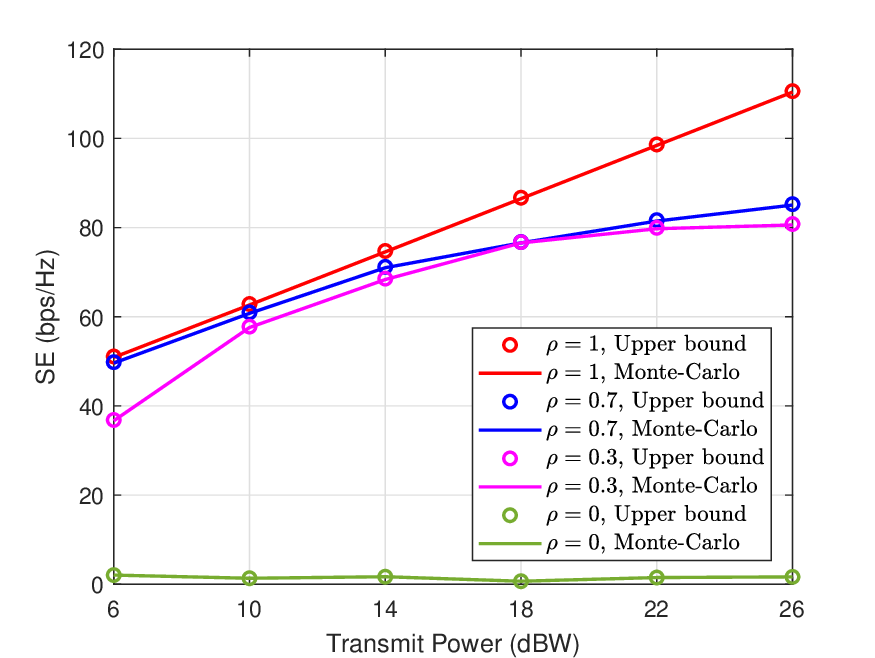}\label{fig_P_Nv18Nh18_se_mcub_rho}}
\hfill
\subfloat[APEB.]{\includegraphics[width=0.4\textwidth]{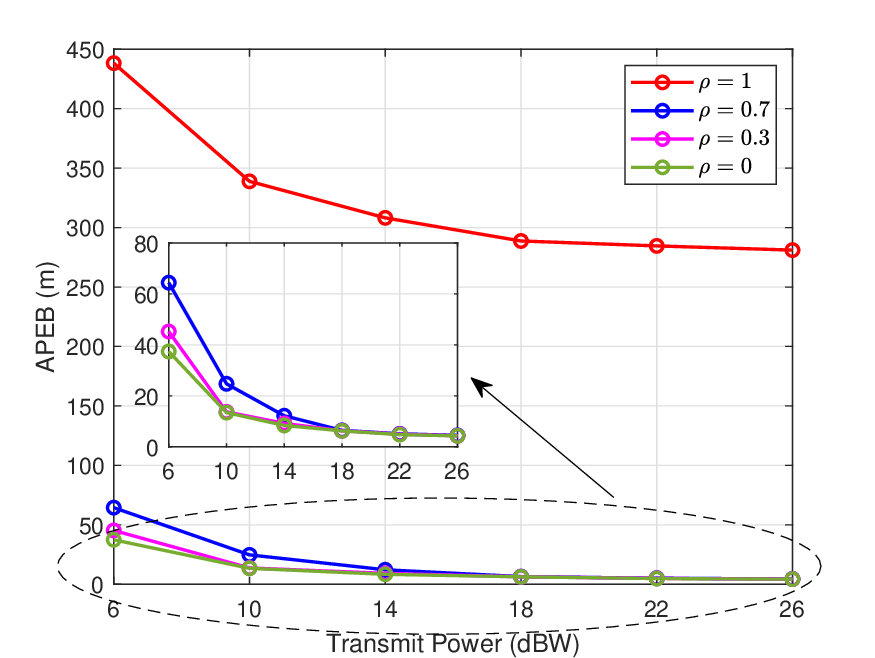}\label{fig_P_Nv18Nh18_speb_rho}}
\caption{SE and APEB performance versus transmit power $P$ with $N_{\rm t}=576$ antennas under the fully connected structure for different weighting coefficients $\rho$.}
\label{fig_P_Nv18Nh18_rho_fc}
\end{figure}

   \begin{figure}[!t]
		\centering
		\includegraphics[width=0.4\textwidth]{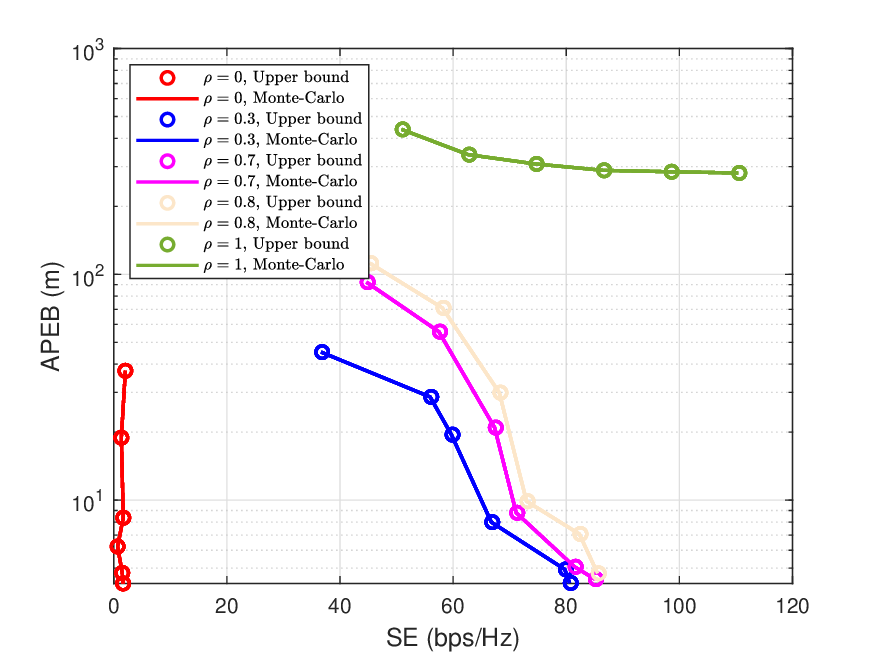}
		\caption{Trade-off between communication SE and localization APEB for different transmit power under typical values of the weighting coefficient.}
        \label{fig:trade_off}
	\end{figure}

\figref{fig_conver} presents the values of the SE and APEB versus the number of iterations under different transmit power budgets with different weighting coefficients for \textbf{Algorithm \ref{alg:algdasam}}.
In general, as the number of iterations increases, the performance of both communications and localization tends to be better.
Then, as depicted in \figref{fig_conver}, the proposed method can converge to certain points for typical scenarios.
\begin{figure}[!t]
\centering
\subfloat[SE.]{\includegraphics[width=0.4\textwidth]{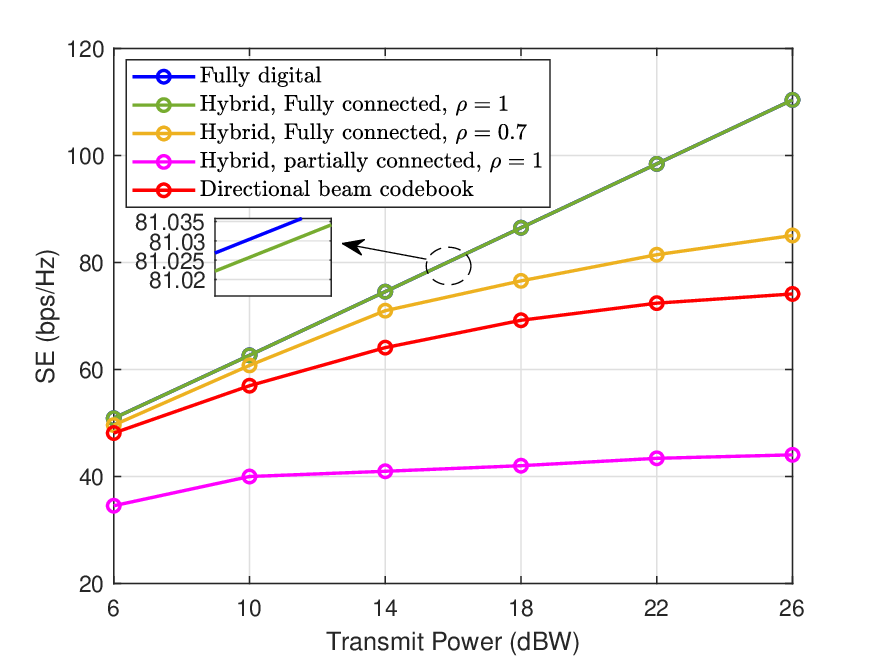}\label{fig_P_Nv18Nh18_se_scheme}}
\hfill
\subfloat[APEB.]{\includegraphics[width=0.4\textwidth]{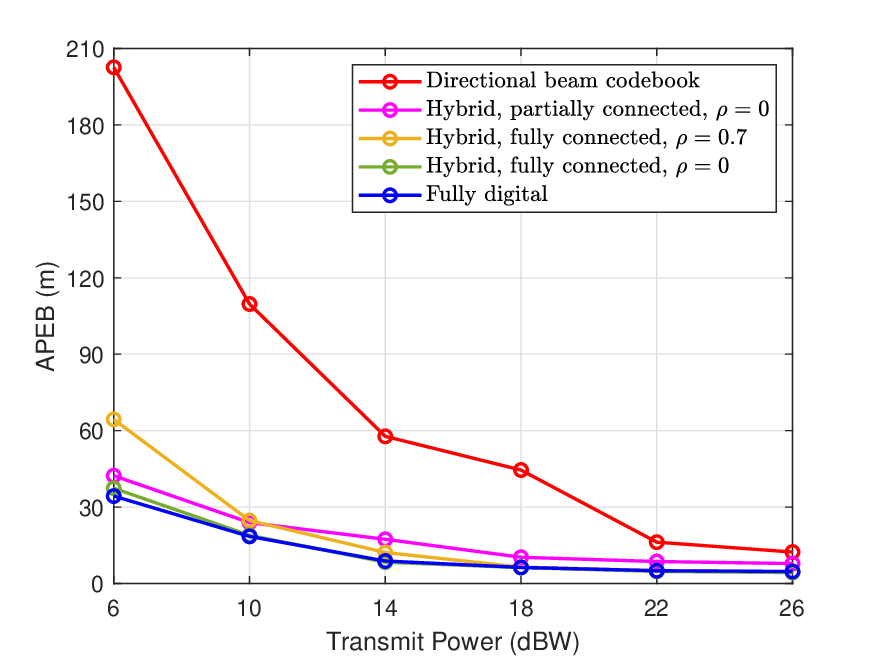}\label{fig_P_Nv18Nh18_scheme}}
\caption{SE and APEB performance versus transmit power $P$ with $N_{\rm t}=576$ antennas for different precoding design strategies.}
\label{fig_P_Nv18Nh18_scheme}
\end{figure}

\figref{fig_P_Nv18Nh18_rho_fc} illustrates the relationship between the transmit power and the SE as well as the APEB.
As observed from \subfigref{fig_P_Nv18Nh18_rho_fc}{fig_P_Nv18Nh18_se_mcub_rho}, the tightness of the upper bound in Eq. \eqref{eq:rateub} is verified.
It is worth noting that $\rho=1$ and $\rho=0$ refer to the pure communication and pure localization system, respectively.
The proposed ICAL system outperforms the pure communication and pure localization ones in APEB and SE, respectively.
Besides, with larger transmit power, the SE sees an increase, and the APEB decreases, leading to better performance of both communications and localization.
Under the scenario that a weighting coefficient $\rho=0.7$ is allocated for communications, both the SE and APEB can be enhanced, especially with higher transmit power.
With a smaller weighting coefficient, i.e., $\rho=0.3$, the ICAL system presents better APEB performance at the cost of the SE performance, compared with the case when $\rho=0.3$.
Moreover, \figref{fig:trade_off} demonstrates the trade-off between communication SE and localization APEB for different transmit power.
Besides, several typical values of the weighting coefficient are considered for their influence on both SE and APEB.
In particular, larger weighting coefficient refers to the scenario that lies emphasis on communications, and thus, as observed, with the same APEB, larger weighting coefficient leads to better SE performance.
The above analysis naturally leads to a conclusion that the proposed ICAL system can offer satisfying communication and localization performance, thanks to the high angular resolution brought by massive MIMO.


\figref{fig_P_Nv18Nh18_scheme} compares the SE and APEB performance with different precoding design strategies as follows:
\begin{itemize}
  \item The hybrid precoders with $N_{\rm rf}$ RF chains under both fully and partially connected structures for different weighting coefficients, as discussed in \secref{sec:prof}.
  \item The optimized fully digital precoder with $N_{\rm t}$ RF chains, i.e., the hybrid precoder with $\bW_{\rm RF}=\bI_{N_{\rm t}}$ and $\bW_{{\rm BB},n}$ optimized only \cite{jeong2018optimization}.
  \item The standard directional beam codebook-based precoder with $N_{\rm t}$ RF chains, i.e., $\bW_n=\sqrt{P/K/N_{\rm sc}}\left[\bv_{1,n}^{\ast},\ldots,\bv_{K,n}^{\ast}\right]$ \cite{keskin2022optimal}.
\end{itemize}
As observed from \figref{fig_P_Nv18Nh18_scheme}, for pure communications and localization, the hybrid scheme with the fully connected structure presents comparable performance with the fully digital scheme with fewer RF chains and thus, less static power consumption.
Note that the proposed ICAL scheme with a fully connected structure for a weighting coefficient, e.g., $\rho=0.7$, is capable for supporting both communications and localization, especially with higher transmit power.
In particular, with 26 dBW transmit power, the proposed ICAL scheme with a fully connected structure can offer the sum SE of approximately 85 bps/Hz for all the UTs, and guarantee the localization precision $< 5$ m for each UT, simultaneously.
Besides, in the low region of the transmit power, the hybrid scheme with a partially connected structure for $\rho=0$ has better APEB performance than with a fully connected one for $\rho=0.7$, at the cost of the SE performance.
In general, the directional beam codebook-based precoder performs worst in the medium and high transmit power regions.
\section{Conclusion}\label{sec:conc}
In this paper, we developed a system that is able to simultaneously perform communications and localization for massive MIMO LEO satellite ICAL systems, by exploiting sCSI.
We derived the performance metrics SE and SPEB for communications and localization, respectively, based on which we formulated a multi-objective problem, to design the hybrid transmitter for the ICAL.
Then, the SE and SPEB maximization problem was converted into the sum-MSE and Euclidean distance minimization one.
We introduced a weighting coefficient to tradeoff between the performance of communications and localization, and formulated an algorithmic framework to handle the corresponding problem.
Simulation results confirmed that the proposed scheme could simultaneously support the communications and localization operations.
Moreover, in this paper, only a single LEO satellite was considered, and future works will focus on the cooperation of multiple satellites to further improve the communication capacity and localization accuracy.
\begin{appendices}
\section{Expression for the
Matrix $\bm{\Xi}_{k}$ in \eqref{eq:trmtr}}\label{app:b}
We denote the rotated position vectors associated with the $k$th UT as
$\bp^{\rm r}_{k}\triangleq \bR\left(\bo\right)^{-1}\left(\bp_{k}-\bq\right)\triangleq\left[p_{k}^{\rm r,x},p_{k}^{\rm r,y},p_{k}^{\rm r,z}\right]^T$
\cite{kwon2021joint},
where $\bR\left(\bo\right)$ is the corresponding rotation matrix, given by \cite{abu2018error,vince2011rotation,chen2022tutorial}
\begin{align}
\bR\left(\bo\right)&\triangleq
\begin{bmatrix}
\cos\varphi_2 &-\sin\varphi_1\sin\varphi_2 &\cos\varphi_1\sin\varphi_2\\
0 &\cos\varphi_1 &\sin\varphi_1\\
-\sin\varphi_2 &-\sin\varphi_1\cos\varphi_2 &\cos\varphi_1\cos\varphi_2
\end{bmatrix}\\
&\triangleq\left[\br_1,\br_2,\br_3\right].
\end{align}
Then, the relationship between the AoD pair and the position of the $k$th UT is given by \cite{abu2018error,vince2011rotation,chen2022tutorial}
\begin{align}
\theta_{k}^{\rm x}&=\arctan\left(\frac{p_{k}^{\rm r,z}}{p_{k}^{\rm r,x}}\right),\
\theta_{k}^{\rm y}=\arccos\left(\frac{p_{k}^{\rm r,y}}{\abs{\abs{\bp_{k}^{\rm r}}}_2}\right).
\end{align}
Besides, the propagation delay and Doppler shifts of LoS path correspond to the position and the velocity information of the $k$th UT as follows \cite{abu2020performance}
\begin{align}
\tau_{k}&=\frac{\abs{\abs{\bp_k-\bq}}_2}{c},\\
\nu_{k}&=-\frac{f_{\rm c}}{c}\frac{\dot{\bp}_k^T\left(\bp_k-\bq\right)}{\abs{\abs{\bp_k-\bq}}_2}.
\end{align}
Then, the component $\bm{\Xi}_{k}$ of the transformation
matrix in \eqref{eq:trmtr} for the $k$th UT can be expressed as
\begin{align}\label{eq:tlos}
\bm{\Xi}_{k}=
\begin{bmatrix}
\frac{\partial \theta_{k}^{\rm x}}{\partial \bp_k} &\frac{\partial \theta_{k}^{\rm y}}{\partial \bp_k} &\frac{\partial \tau_{k}}{\partial \bp_k} &\frac{\partial \nu_{k}}{\partial \bp_k} \vspace{1ex}
\end{bmatrix}
\in \mathbb{R}^{3\times 4},
\end{align}
where the elements are detailed as
\begin{subequations}
\begin{align}
\frac{\partial \theta_{k}^{\rm x}}{\partial \bp_k}&=\frac{p_k^{\rm r,x}\br_3-p_k^{\rm r,z}\br_1}{\left(p_k^{\rm r,x}\right)^2+\left(p_k^{\rm r,z}\right)^2},\\
\frac{\partial \theta_{k}^{\rm y}}{\partial \bp_k}&=\frac{p_k^{\rm r,y}\left(\bp_k-\bq\right)-\br_2\abs{\abs{\bp_k^{\rm r}}}_2^2}{\abs{\abs{\bp_k^{\rm r}}}_2^2\sqrt{\left(p_k^{\rm r,x}\right)^2+\left(p_k^{\rm r,z}\right)^2}},\\
\frac{\partial \tau_{k}}{\partial \bp_k}&=\frac{1}{c}\frac{\bp_k-\bq}{\abs{\abs{\bp_k-\bq}}_2},\\
 \frac{\partial \nu_{k}}{\partial \bp_k}&=\frac{f_{\rm c}}{c}\frac{\left(\dot{\bp}_k^T\left(\bp_k-\bq\right)\right)\left(\bp_k-\bq\right)-\abs{\abs{\bp_k-\bq}}_2^2\dot{\bp}_k}{\abs{\abs{\bp_k-\bq}}_2^3}.
\end{align}
\end{subequations}

\section{Proof for Eq. \eqref{eq:optC}}\label{app:g}
The precoder $\bW_n^{\rm loc}$ can be decomposed as
\begin{align}
\bW_n^{\rm loc}=\bm{\Pi}_{\bL_n}\bW_n^{\rm loc}+\bm{\Pi}_{\bL_n}^{\bot}\bW_n^{\rm loc},
\end{align}
where $\bm{\Pi}_{\bL_n}\triangleq \bL_n\left(\bL_n^H \bL_n\right)^{-1}\bL_n^H$ denotes the subspace spanned by the columns of $\bL_n$ and $\bm{\Pi}_{\bL_n}^{\bot}\triangleq\bI_{N_{\rm t}}-\bm{\Pi}_{\bL_n}$.
Following the equality that $\bC_n=\bW_n^{\rm loc}\left(\bW_n^{\rm loc}\right)^H$, we have
\begin{align}
\bC_n&=\bW_n^{\rm loc}\left(\bW_n^{\rm loc}\right)^H=\bm{\Pi}_{\bL_n}\bW_n^{\rm loc}\left(\bW_n^{\rm loc}\right)^H\left(\bm{\Pi}_{\bL_n}\right)^H+\tilde{\bC}_n,
\end{align}
It is worth noting that since $\bv_{k,n}$, $\bv_{k,n,\theta_k^{\rm x}}$ and $\bv_{k,n,\theta_k^{\rm y}}$ belong to the subspace $\bm{\Pi}_{\bL_n}$, it is not difficult to verify that the terms $\bv_{k,n}^H\tilde{\bC}_n\bv_{k,n}$, $\bv_{k,n}^H\tilde{\bC}_n\bv_{k,n,\theta_k^{\rm x}}$, $\bv_{k,n}^H\tilde{\bC}_n\bv_{k,n,\theta_k^{\rm y}}$, $\bv_{k,n,\theta_k^{\rm x}}^H\tilde{\bC}_n\bv_{k,n,\theta_k^{\rm x}}$, $\bv_{k,n,\theta_k^{\rm y}}^H\tilde{\bC}_n\bv_{k,n,\theta_k^{\rm y}}$, $\bv_{k,n,\theta_k^{\rm x}}^H\tilde{\bC}_n\bv_{k,n,\theta_k^{\rm y}}$ are all equal to zeros.
Thus, it can be concluded that the FIM does not depend on $\tilde{\bC}_n$.
In addition, we have
\begin{align}\label{eq:appg4}
{\rm Tr}\left(\tilde{\bC}_n\right)&={\rm Tr}\left(\bm{\Pi}_{\bL_n}\bW_n^{\rm loc}\left(\bW_n^{\rm loc}\right)^H\bm{\Pi}_{\bL_n}^{\bot}\right.\notag\\
&\qquad \qquad\left.+\bm{\Pi}_{\bL_n}^{\bot}\bW_n^{\rm loc}\left(\bW_n^{\rm loc}\right)^H\bm{\Pi}_{\bL_n}\right.\notag\\
&\qquad \qquad \qquad \left.+\bm{\Pi}_{\bL_n}^{\bot}\bW_n^{\rm loc}\left(\bW_n^{\rm loc}\right)^H\bm{\Pi}_{\bL_n}^{\bot}\right)\notag\\
&=\abs{\abs{\left(\bW_n^{\rm loc}\right)^H\bm{\Pi}_{\bL_n}^{\bot}}}_F^2\geq 0.
\end{align}
Furthermore, $\bC_n$ is subject to the constraint in Eq. \eqref{eq:spebmin3b}, and thus, it can be concluded that ${\rm Tr}\left(\tilde{\bC}_n\right)=0$, otherwise the constraint in Eq. \eqref{eq:spebmin3b} is not satisfied.
As observed from Eq. \eqref{eq:appg4}, ${\rm Tr}\left(\tilde{\bC}_n\right)=0$ is equivalent to $\left(\bW_n^{\rm loc}\right)^H\bm{\Pi}_{\bL_n}^{\bot}=0$, i.e., $\tilde{\bC}_n=0$.
Therefore,
\begin{align}
\bC_n=\bm{\Pi}_{\bL_n}\bW_n^{\rm loc}\left(\bW_n^{\rm loc}\right)^H\left(\bm{\Pi}_{\bL_n}\right)^H\triangleq \bL_n\bZ_n\bL_n^H,
\end{align}
where $\bZ_n=\left(\bL_n^H \bL_n\right)^{-1}\bL_n^H\bW_n^{\rm loc}\left(\bW_n^{\rm loc}\right)^H\bL_n\left(\bL_n^H \bL_n\right)^{-1}$.
\section{Steps for Handling Eqs. \eqref{eq:bcd1} -- \eqref{eq:updss}}\label{app:a}
The steps for handling Eqs. \eqref{eq:bcd1} -- \eqref{eq:updss} are presented as follows:
\subsubsection{Update of $\bG_{k,n}$}
We formulate the optimization problem to update $\bG_{k,n}$ as
    \begin{subequations}\label{eq:wscl4}
    \begin{align}
    \mathcal{Q}_4:&\mathop{\mathrm{minimize}}\limits_{\mathcal{G}}\ \ f\left(\mathcal{G}\right) \label{eq:wscl4a}\\
    &\qquad \ \ \mathrm{s.t.}\ \  \sum_{k=1}^K\sum_{n=1}^{N_{\rm sc}}{\rm Tr}\left\{\bG_{k,n}\bG_{k,n}^H\right\}\leq PK. \label{eq:wscl4b}
    \end{align}
    \end{subequations}
To handle problem $\mathcal{Q}_4$, we introduce a  Lagrange multiplier for constraint in \eqref{eq:wscl4b} and the corresponding Lagrange function for $\mathcal{Q}_4$ is given by \cite{boyd2004convex}
\begin{align}
f\left(\mathcal{G},\mu\right)&=\sum_{k=1}^K\sum_{n=1}^{N_{\rm sc}}L\left(\bG_{k,n}\right)\notag\\
&\qquad +\mu\left(\sum_{k=1}^K\sum_{n=1}^{N_{\rm sc}}{\rm Tr}\left\{\bG_{k,n}\bG_{k,n}^H\right\}-PK\right).
\end{align}
Subsequently, by utilizing the Karush-Kuhn-Tucker (KKT) conditions of $\mathcal{Q}_4$, $\bG_{k,n}$ can be updated by $\bG_{k,n}=\left(\bA_{k,n}+\mu\bI_{N_{\rm t}}\right)^{-1}\bm{\Psi}_{k,n}$,
where $\bA_{k,n}$ is a Hermitian matrix which can be decomposed as $\bA_{k,n}=\bD_{k,n}\bm{\Lambda}_{k,n}\bD_{k,n}^H$ and the expression for $\bA_{k,n}$, $\bm{\Psi}_{k,n}$ is given by
\begin{align}
\bA_{k,n}&=\rho\omega_{k,n}\abs{u_{k,n}}^2\bar{\bh}_{k,n}^{\ast}\bar{\bh}_{k,n}^T+\left(\frac{1-\rho}{K}+\frac{\eta_{k,n}}{2}\right)\bI_{N_{\rm t}},\\
\bm{\Psi}_{k,n}&=\frac{1-\rho}{K}\bW_n^{\rm loc}+\rho\omega_{k,n}u_{k,n}^{\ast}\bar{\bh}_{k,n}^{\ast}\be_k^T\notag\\
&\quad \quad \quad \quad \quad +\frac{\eta_{k,n}}{2}\left(\bW_{\rm RF}\mathbf{W}_{{\rm BB},n}-\bQ_{k,n}\right).
\end{align}
Then, the left hand side of Eq. \eqref{eq:wscl4b} can be written as a function of $\mu$, which is given by
\begin{align}
\delta\left(\mu\right)&=\sum_{k=1}^K\sum_{n=1}^{N_{\rm sc}}{\rm Tr}\left\{\bG_{k,n}\left(\mu\right)\bG_{k,n}^H\left(\mu\right)\right\}\notag\\
&=\sum_{k=1}^K\sum_{n=1}^{N_{\rm sc}}\sum_{i=1}^{N_{\rm t}}\frac{\left[\bm{\Phi}_{k,n}\right]_{i,i}}{\left(\left[\bm{\Lambda}_{k,n}\right]_{i,i}+\mu\right)^2},
\end{align}
where $\bm{\Phi}_{k,n}=\bD_{k,n}^H\bm{\Psi}_{k,n}\bm{\Psi}_{k,n}^H\bD_{k,n}$.
The Lagrange multiplier $\mu$ can be found to satisfy the complementarity slackness condition \cite{boyd2004convex}.
In particular, if $\delta\left(0\right)\leq PK$, then $\mu=0$; otherwise $\mu$ is selected to satisfy $\delta\left(\mu\right)=PK$ via the bisection search.
\subsubsection{Update of $u_{k,n}$}
The update for $u_{k,n}$ is
$u_{k,n} = \be_k^T\bG_{k,n}^H\bar{\bh}_{k,n}^{\ast}(\sum_{i=1}^K|\bar{\bh}_{k,n}^T\bG_{k,n}\be_i|^2+N_0)^{-1}$.
\subsubsection{Update of $\omega_{k,n}$}
The update for $\omega_{k,n}$ is given by $\omega_{k,n}=(|u_{k,n}^{\ast}\bar{\bh}_{k,n}^T\bG_{k,n}\be_k-1|^2+\sum_{i\neq k}|u_{k,n}^{\ast}\bar{\bh}_{k,n}^T\bG_{k,n}\be_i|^2+N_0|u_{k,n}|^2)^{-1},\ \forall k,n$.
\subsubsection{Update of $\bW_{{\rm BB},n}$}
The optimization problem to update $\bW_{{\rm BB},n}$ is formulated as
    \begin{align}\label{eq:wscl5}
    \mathcal{Q}_5:&\mathop{\mathrm{maximize}}\limits_{\{\mathbf{W}_{{\rm BB},n}\}_{n=1}^{N_{\rm sc}}}\ \ f\left(\bar{\mathbf{W}}_{\rm BB}\right)=\sum_{k=1}^K\sum_{n=1}^{N_{\rm sc}} \frac{\eta_{k,n}}{2}\abs{\abs{\bG_{k,n}\right.\right.\notag\\
    &\qquad \qquad \qquad \qquad \left.\left.-\bW_{\rm RF}\mathbf{W}_{{\rm BB},n}+\bQ_{k,n}}}_F^2. 
    \end{align}
For both fully and partially connected structures, we set the derivative of the objective function $f\left(\bar{\mathbf{W}}_{\rm BB}\right)$ with respect to $\bW_{{\rm BB},i}$ to be zero,
and then, the expression to update $\bW_{{\rm BB},i}$ is given by $\mathbf{W}_{{\rm BB},i}=\frac{1}{\sum_{k=1}^K\eta_{k,i}}\left(\bW_{\rm RF}^H\bW_{\rm RF}\right)^{-1}\sum_{k=1}^K\eta_{k,i}\bW_{\rm RF}^H\left(\bG_{k,i}+\bQ_{k,i}\right)$.
In particular, for partially connected structure, we have $\bW_{\rm RF}^H\bW_{\rm RF}=N_{\rm g}\bI_{\rm N_{\rm rf}}$.
\subsubsection{Update of $\bW_{\rm RF}$}
The optimization problem to update $\bW_{\rm RF}$ is formulated as
    \begin{subequations}\label{eq:wscl5}
    \begin{align}
    \mathcal{Q}_6:&\mathop{\mathrm{maximize}}\limits_{\bW_{\rm RF}}\ \ \sum_{k=1}^K\sum_{n=1}^{N_{\rm sc}} \frac{\eta_{k,n}}{2}\abs{\abs{\bG_{k,n}-\bW_{\rm RF}\mathbf{W}_{{\rm BB},n}\right.\right.\notag\\
    &\qquad \qquad \qquad \qquad \qquad \qquad \qquad \left.\left.+\bQ_{k,n}}}_F^2 \label{eq:wscl6a}\\
    &\qquad \ \ \ \mathrm{s.t.}\ \ \bW_{\rm RF}\in \mathcal{S}. \label{eq:wscl6b}
    \end{align}
    \end{subequations}
For the fully connected structure, $\bW_{\rm RF}$ can be updated as $\bW_{\rm RF}=\exp\left\{-\jmath\angle \bX^T\right\}$ \cite{arora2019hybrid},
where $\bX=\sum_{k=1}^K\sum_{n=1}^{N_{\rm sc}}\frac{\eta_{k,n}}{2}\bW_{{\rm BB},n}\left(\bG_{k,n}+\bQ_{k,n}\right)^H-(\bT-\lambda_{\rm max}(\bT)\bI_{N_{\rm rf}})\bW_{\rm RF}^H$
and $\lambda_{\rm max}(\bT)$ denotes the maximum eigenvalue of $\bT=\sum_{k=1}^K\sum_{n=1}^{N_{\rm sc}}\frac{\eta_{k,n}}{2}\bW_{{\rm BB},n}\bW_{{\rm BB},n}^H$ \cite{arora2019hybrid}.
For the partially connected structure, by expanding the objective in \eqref{eq:wscl6a} and utilizing the  property that $\bW_{\rm RF}^H\bW_{\rm RF}=N_{\rm g}\bI_{\rm N_{\rm rf}}$, the problem can be rewritten as \cite{yu2016alternating}
    \begin{align}
    \mathcal{Q}_7:&\mathop{\mathrm{maximize}}\limits_{\left[\bW_{\rm RF}\right]_{i,j}}\ \ \Re\left\{{\rm Tr}\left\{\bY_{i,j}^H\left[\bW_{\rm RF}\right]_{i,j}^{\ast}\right\}\right\} \label{eq:wscl7a}\\
    &\qquad \ \ \ \mathrm{s.t.}\ \ \abs{\left[\bW_{\rm RF}\right]_{i,j}}=1, \ \forall i, \ \forall j=\left\lceil \frac{i}{N_{\rm g}}\right\rceil, \label{eq:wscl7b}
    \end{align}
where $\bY_{i,j}=\sum_{k=1}^K\sum_{n=1}^{N_{\rm sc}} \frac{\eta_{k,n}}{2}\left[\bG_{k,n}+\bQ_{k,n}\right]_{i,:}\left[\mathbf{W}_{{\rm BB},n}\right]_{j,:}^H$.
Then, the $(i,j)$th element of $\bW_{\rm RF}$ can be updated by $\left[\bW_{\rm RF}\right]_{i,j}=\exp\left\{\angle \bY_{i,j}\right\}, \ \forall i, \ \forall j=\left\lceil i/N_{\rm g}\right\rceil$.
\end{appendices}

\bibliographystyle{IEEEtran}
\bibliography{Refabrv_20180802,References_20211203,References_20230628}

\end{document}